\documentclass[]{spie}  

\usepackage{amsmath,amsfonts,amssymb}
\usepackage{graphicx}
\usepackage[colorlinks=true, allcolors=blue]{hyperref}

\usepackage[a4paper,
            bindingoffset=0.2in,
            left=0.875in,
            right=0.875in,
            top=1.75in,
            bottom=1.25in,
            footskip=.25in]{geometry}

\title{Modal Simulation Framework for the Design and Verification of Future Few-Mode Far-Infrared Spectrometers}

\author[a,b]{Bram N.R. Lap}
\author[a,b]{Willem Jellema}
\author[c]{Stafford Withington}
\author[d]{David A. Naylor}

\affil[a]{Kapteyn Astronomical Institute, University of Groningen, 9700 AV, Groningen, The Netherlands}
\affil[b]{SRON Netherlands Institute for Space Research, 9700 AV, Groningen, The Netherlands}
\affil[c]{Cavendish laboratory, JJ Thomson Avenue, Cambridge CB3 OHE, UK}
\affil[d]{Institute for Space Imaging Science, Department of Physics and Astronomy, University of Lethbridge, 4401 University Drive, Lethbridge, Alberta T1K 3M4, Canada}

\authorinfo{Further author information: (Send correspondence to B.Lap.)\\B.N.R.L: E-mail: b.lap@sron.nl, Telephone: +31 (0)6 26 78 55 65\\  W.J: E-mail: w.jellema@sron.nl, Telephone: +31 (0)6 98 76 54 32 \\ S.W.: E-mail: stafford@mrao.cam.ac.uk, Telephone: Telephone: +44 (0)1223 337393\\ D.A.N.: naylor@uleth.ca, Telephone: 	+1 403 329 2426}

\begin{document} 
\maketitle

\begin{abstract}
The next generation of astronomical space-based far-infrared (FIR) missions require ultra-sensitive spectroscopy as a diagnostic tool. These instruments use ultra-sensitive detector technologies to attain unprecedented levels of spectral observing sensitivity. The reception patterns of the individual detectors consist of individually coherent orthogonal field distributions, or equivalently, they are few-mode (5-20), to increase the spectral-spatial coupling to the astronomical source. However, the disadvantage of few-mode detectors is an increase in coupling to external (from the sky or warm telescope optics) and internal (from the instrument itself) straylight, which can greatly affect the measurement of the source spectrum. Therefore, understanding the spectral-spatial few-mode behaviour of these systems in detail, and developing verification and calibration strategies, are crucial to ensure that the science goals of these future mission are met. 
Since conventional modelling techniques are less suited to address this problem, we developed a modal framework to model, analyse, and address these issues \cite{Lap:22}. In this paper, we use \textit{Herschel}'s Spectral and Photometric Imaging REceiver (SPIRE) as a case study, because its optical design is representative for future FIR missions and illustrative to highlight calibration issues observed in-flight, while including straylight. Our analysis consist out of two part. In the first part, we use our modal framework to simulate the few-mode SPIRE Fourier Transform Spectrometer (FTS). In the second part, we carry out a end-to-end frequency-dependent partially coherent analysis of \textit{Herschel}-SPIRE. These simulations offer a qualitative explanation for the few-mode behaviour observed in-flight. Furthermore, we use the \textit{Herschel}-SPIRE case-study to demonstrate how the modelling framework can be used to support the design, verification and calibration of spectrometers for future FIR missions. The modal framework is not only limited to the spectrometers discussed, but it can be used to simulated a wide range of spectrometers, such as low-resolution gratings and high-spectral resolution Fabry-Pérot interferometers.  
\end{abstract}


\section{INTRODUCTION}
\label{sec:intro}  

In FIR astronomy (FIR, $30 – 300 \mu\text{m}$) the majority of the science questions focus on the formation and evolution of celestial objects such as galaxies, stars, and planets in extrasolar planetary systems \cite{Duncan:19,kamp:21,Roelfsema:2018}. Broadband spectroscopy is the primary diagnostic tool to constrain the evolution models of these astronomical objects, because the instrument spatial resolution is prohibited by the cost of the required cold, monolithic, primary mirrors. Although limited in spatial resolution, the next generation of FIR astronomical instruments can achieve unprecedented levels of spectral observing sensitivities by using ultra-sensitive large-format detector technologies.

To increase the optical throughput of these ultra-sensitive systems, partially coherent, or equivalently few-mode detector schemes are used, i.e. the reception pattern of an individual detector consist of individually coherent, but mutually incoherent field distributions. An example of this are absorber coupled detectors, opposed to detectors operating in single-mode limit. However, the disadvantage of these few-mode detectors, is that they produce a few-mode on-sky instrument beam pattern, opposed to single-mode systems, and they are intrinsically prone to increased straylight coupling. Both mechanisms can greatly affect the instrument calibration and the scientific observations. Consequently, there is a strong need to understand the few-mode behaviour of these instruments in detail, to control their response to straylight, and to develop adequate verification and calibration strategies to ensure that the scientific performance of the instrument can be met. In other words, end-to-end few-mode optical system modelling is required. However, classical modelling techniques can not account for few-mode partial coherence and associated behaviour of complex optical systems. In an earlier paper we developed a modal modelling framework to address this problem \cite{Lap:22}. In this paper, we build upon this work to i) demonstrate how this modelling technique can be used for the end-to-end frequency-dependent modelling of complex few-mode optics, and ii) to highlight how this technique can be to develop adequate design, verification, and calibration strategies for future FIR missions. 

We use \textit{Herschel}-SPIRE as a case study, because its optical design is representative for future FIR missions, and illustrative to highlight calibration issues observed in-flight \cite{Makiwa:13}. Before this system level behaviour of \textit{Herschel}-SPIRE can be studied, the partially coherent few-mode behaviour of its FTS ($R\sim$1000) must be first understood. In this paper, we will use our modal framework to study the few-mode \textit{Herschel}-SPIRE instrument, which is typified by the FTS. To our knowledge, such a partially coherent analysis of a few-mode FTS has not yet been done before.
Next, we turn to the end-to-end frequency-dependent partially coherent modelling of \text{Herschel}-SPIRE, the in-flight calibration of which revealed clearly that it experienced few-mode behaviour. For example, \textit{Herschel}-SPIRE was properly calibrated for the two extremes spatial extend at which astronomical source appear on the sky, point and extended sources. However, the majority of astronomical objects are semi-extended, and the in-flight calibration for such sources was problematic \cite{Swinyard:14,Valtchanov:17}. It was hypothesized that the spatial extend of the source combined with the few-mode detectors caused the system to experience partially coherent behaviour. Our aim is to confirm this hypothesis by performing an end-to-end partially coherent analysis of \textit{Herschel}-SPIRE. In these simulations, we include the spectral and spatial distribution of the source, the entire \textit{Herschel}-SPIRE optical train, which was also few-mode, and the partially coherent nature of the detectors. In addition, we will use this end-to-end \textit{Herschel}-SPIRE simulator to demonstrate how the framework can be used to support the design, verification, and calibration of future spectroscopic FIR missions, in the presence of external and internal straylight.

\section{THEORY}
\label{sec:theory}

SPIRE is an imaging submillimetre spectrometer on-board the \text{Herschel} Space Observatory. This instrument comprises a three-band imaging photometer, and a two-band imaging Fourier Transform operating from $447–1550\,$GHz \cite{Griffin:10}. In this paper, we will focus on the SPIRE Long Wavelength (SLW) band of the \textit{Herschel}-SPIRE spectrometer, because the partially coherent behaviour was particularly pronounced for this case \cite{Makiwa:13}. From now on, we will refer to the SLW band of the \textit{Herschel}-SPIRE spectrometer as \textit{Herschel}-SPIRE. 

 

\subsection{Optical Model}

The \textit{Herschel}-SPIRE optical train consists of five subsystems \cite{Griffin:03}: i) the telescope, ii) the common fore optics, iii) the 2nd relay, iv) the FTS, and v) the camera (bay) optics. From now on the telescope subsystem and the collection of optical surfaces from the telescope focal plane to the detector plane (i.e. excluding the detector) will be referred to as the\text{Herschel} telescope and the \text{Herschel}-SPIRE instrument, respectively.

The \text{Herschel} telescope couples the incident spectral electromagnetic (EM) field from a distant astronomical object to the telescope focal plane. The spectral and spatial distribution of the astronomical source can follow any arbitrary form, where the former is referred to as the source spectrum. The common-fore optics takes the image of the object at the telescope focal plane and presents it, with the correct focal ratio, to the \textit{Herschel}-SPIRE spectrometer. The second optical relay is needed to bridge the distance from the focal plane of the common fore optics to the input of the FTS. 
\textit{Herschel}-SPIRE uses a Mach-Zehnder FTS \cite{Mach:92,Zehnder:91}, but in this paper we will use a Michelson interferometer \cite{Michelson:90} instead, because the implementation of the Michelson is simpler, and its operation principles are still representative for FTSs in general. In an FTS, the interference between the two arms is determined by the Optical Path Difference (OPD) between the two arms. There is constructive interference when there is Zero Optical Path (ZPD) difference (the movable mirrors is in its reference position), but when a nonzero OPD is introduced (the movable mirror is not in its reference position) the interference between the two beams changes. The power in this interference pattern is measured as function as OPD, which is referred to as the interferogram. In \textit{Herschel}-SPIRE, the interferogram is measured by the last subsystem, i.e. camera (bay) optics. This subsystem focuses the output image plane of the FTS onto the detector focal plane via an intermediate detector pupil. At the detector plane, a detector array of feedhorn-coupled bolometers is placed, i.e. the spectrometer is an imaging FTS. The detectors measure the optical power in the EM field over the detector plane as function of OPD, i.e. for each detector in the array an interferogram is measured. Then, by taking the Fourier Transform of these interferograms the measured source spectrum is obtained. Finally, the measured spectrum is calibrated, and ready to be analysed by astronomers. 

Before we can model \textit{Herschel}-SPIRE, we need to explain the underlying assumptions of the simulations. These assumptions are discussed in detail in Ref.~\citenum{Lap:22}, but summarized here. First, we assume that \textit{Herschel}-SPIRE is subject to linear physical optics, and we will ignore polarization effects: we will consider linear polarized electric fields. Second, the optical elements are optically thin, therefore refraction and reflection effects within the optics are ignored, which is applicable at FIR wavelengths \cite{Goldsmith:98}. Third, the polychromatic partially coherent behaviour of \textit{Herschel}-SPIRE can be described by a collection of mutually incoherent monochromatic realizations. The latter is valid when the random fluctuations in the electric fields are narrow in bandwidth $\Delta \nu$ compared to the mean frequency $\bar{\nu}$ \cite{Wolf:07}, and when complex statistically stationary electric fields are considered, both of these conditions are met in \textit{Herschel}-SPIRE.

\begin{figure} [b]
   \begin{center}
   \begin{tabular}{c} 
   \includegraphics[width=.9\textwidth]{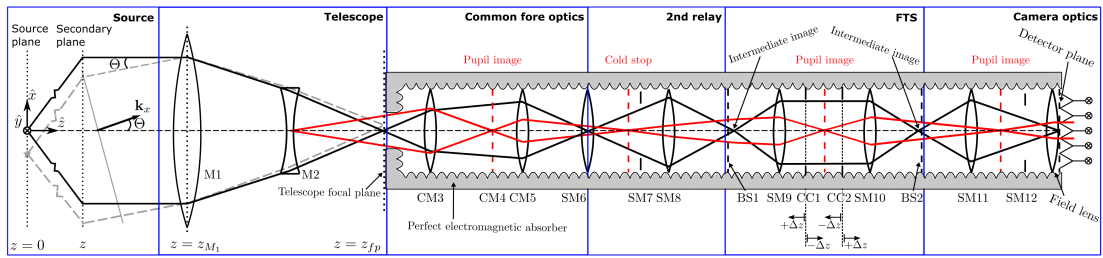}
   \end{tabular}
   \end{center}
   \caption[example] 
   {\label{fig:1} 
1-D inline equivalent optical model of \textit{Herschel}-SPIRE. The spectral light originating from a point source enters the telescope under a certain angle $\Theta$ from the left-hand side. The telescope subsystem creates an image on the telescope focal plane. The common fore optics relays this image to the input of the SPIRE instrument, the $2$-nd relay is used to feed the spectral field to the spectrometer, and its FTS. The camera optics focuses the spectral light onto the detector plane, where a detector array is place to measure the power in the field as function of OPD.}
\end{figure} 

We developed the Huygens-Fresnel Modal Framework (HFMF) to enable the partially coherent modelling of few-mode optics \cite{Lap:22}. Although three-dimensional (3-D) optical simulations have been carried out, for now we will consider one-dimension (1-D) optics. This assumptions, without the loss generality, simplifies the modelling considerably, and allows us to gain a conceptual appreciation of the operation principles of complex few-mode optics, such as the case-study discussed in this paper. 

We construct an 1-D inline equivalent model of \textit{Herschel}-SPIRE using the SPIRE design description \cite{Griffin:03}. The calculated geometrical and optical parameters are derived and verified using the lensmaker's formula \cite{Hecht:99}, literature values \cite{Sein:03}, and an optical software package \footnote{ (2022) Zemax (OpticsStudio). \url{https://www.zemax.com/pages/opticstudio} }. The optical software allows us to trace the object and the pupil through the system, which we call “object imaging” and “pupil imaging” mode, respectively. In the object imaging mode, we trace the light path of the incident light using ray tracing. In the pupil imaging mode, we ray trace the pupil stop, which is positioned at the telescope secondary mirror (M2), through the optics, and determines the locations of the pupils. The resulting 1-D \textit{Herschel}-SPIRE optical model is shown in Fig. \ref{fig:1}, where the details on the optical elements are listed in Table \ref{tab:1}. In Fig. \ref{fig:1}, the $\hat{z}$ axis is defined along the optical axis (the $\hat{z}$ axis), which coincides with transverse direction of the propagating electric field, the $\hat{x}$ axis is defined perpendicular to $\hat{z}$, and $\hat{y}$ axis points out of the paper. Moreover, the 1-D in-line optically thin optical elements, such as apertures and thin-lenses, replacing the 3-D optical elements, are defined in the $xz$ plane, and their optical surfaces are defined perpendicular to the $\hat{z}$ axis. Here, the 1-D optical elements extend to infinity along $\hat{y}$, to avoid any spatial filtering effects in this direction. Finally, $\Delta z$ shows how we introduce an OPD into the FTS. Here, an OPD of $+ 2 \Delta z$ and $- 2 \Delta z$ can be seen as lengthening or shortening the propagation distance from the beamsplitter (BS1), to the movable mirror, and back to the beamsplitter (B21), respectively.

In the following, we provide a mathematical description of our numerical modelling scheme for monochromatic light at a single frequency $\nu$. We will first describe the telescope system (the first subsystem) and how it couples to the incident electric field from an on-sky point source using continuous forms. Next, we convert the continuous description of the telescope system into a discrete matrix representation, to allow for the subsystem to be numerically modelled. Finally, the numerical modelling of the last four subsystems (the common fore optics, the 2nd relay, the FTS, and the camera optics), i.e. the \textit{Herschel}-SPIRE instrument, using the HFMF is discussed. Later, we combine the monochromatic results to provide a polychromatic partially coherent description of \textit{Herschel}-SPIRE.

\subsection{Telescope subsystem}
\label{sec:theory:telescope_subsystem}


The 1-D mathematical description of the telescope subsystem provided in this section follows the derivation in 2-D given by Ref.~\citenum{Chen:18}. We start with considering a point source in the far-field of the telescope, the spectral content of which is given in spectral power per unit bandwidth per unit area ($\text{WHz}^{-1}\text{m}^{-2}$) also called flux density. The electric field originating from this point source creates a 1-D secondary field, $\boldsymbol{e}(\rm{\textbf{x}})$, which is fully coherent. A single realization of this 1-D electric field at frequency $\nu$ can be written as a linear combination of plane waves \cite{Orfanidis:16}
\begin{equation}
    \boldsymbol{e}( \textbf{x}, z ) = \frac{e_0}{ 2 \pi} \int \text{exp}( i \textbf{k}_x \cdot \textbf{x} ) \text{exp}(i k z ) d {k_x}
    \label{eq:3}
\end{equation}
where $e_0$ is the complex amplitude, $k_x$ is the wave vector along the $\hat{x}$ axis, $\textbf{x}$ is the position vector, and $z$ is the propagation distance. We consider the time-dependent phase term $ \text{exp}(-iwt)$, and for \textit{Herschel}-SPIRE we can assume paraxial optics
\begin{align}
    \textbf{k}_x & \approx k \boldsymbol{\Theta}, \label{eq:4} \\
    k_z &\approx k, 
    \label{eq:5}
\end{align}
where $k = 2 \pi / \lambda$, and $\Theta$ is the angle between the transverse direction of the incident secondary field $\boldsymbol{e}$ and the optical axis (see Fig. \ref{fig:1}). Moreover, from Eq. (\ref{eq:4}) we get $ d k_x \approx k d \Theta$, and a single coherent plane wave at the telescope primary (M1) is given by
\begin{equation}
    \boldsymbol{e}_{M_1} ( \boldsymbol{x}_{M_1} , z_{M_1} ) = \frac{k e_0 }{2 \pi}  \text{exp}(i k z_{M_1} ) \text{exp}(i k \boldsymbol{\Theta} \cdot \boldsymbol{\rm{x}}_{M_1} ) d \Theta.
    \label{eq:6}
\end{equation}
where $z_{M_1}$ is the position of M1, and $\boldsymbol{\rm{x}}_{M_1}$ is the position vector describing the surface of M1.

In accordance with the paraxial and optical thin approximation, the primary mirror can be represented by a thin lens \cite{Goodman:05}. Therefore, upon passing through the primary mirror, the electric field is truncated by the lens, which is in 1-D is represented by a box-car function, $ \rm{ \boldsymbol{\Pi} (D_{M_1}) } $, where $\rm{D_{M_1}}$ is the diameter of the primary. Moreover, a quadratic phase factor is accumulated. The resulting field becomes 
\begin{equation}
    \boldsymbol{e}_{M_1} ( \boldsymbol{\rm{x}}_{M_1} , z_{M_1} ) = \frac{k e_0}{ 2 \pi } exp(i k z_{M_1} ) exp(i k \boldsymbol{\rm{\Theta}} \cdot \boldsymbol{\rm{x}}_{M_1} ) \rm{ \boldsymbol{\Pi} }(D_{M_1}) exp \Big( \frac{-i k \boldsymbol{\rm{x}}_{M_1}^2}{2f} \Big) d \Theta,
    \label{eq:7}
\end{equation}
where $f$ is the focal length of the primary, and $\text{exp} [ -i k \boldsymbol{\rm{x}}_{M_1}^2/(2f) ] $ is the quadratic phase factor \cite{Goodman:05}. In Eq. (\ref{eq:7}), the flux density from the source is converted into spectral power per unit band width ($\text{WHz}^{-1}$), i.e. a Power Spectral Density (PSD).

Scalar diffraction theory, i.e. Fraunhofer diffraction \cite{Goodman:05}, can be used to map the field over the primary mirror, through the secondary mirror (M2), onto the telescope focal plane, and Eq. (\ref{eq:7}) can be written into  
\begin{equation}
    \boldsymbol{e}_{fp} ( \rm{\textbf{x}}_{fp}, z_{fp} ) = \frac{1}{\lambda^2 F\#} \Big[ \text{sinc}(u) - \epsilon \text{sinc}(u) \Big] d \Theta.
    \label{eq:8}
\end{equation}
Here, $_{fp}$ labels the telescope focal plane, $\rm{\textbf{x}}_{fp}$ is the position vector in the telescope focal plane, $\rm{z_{fp}}$ is the position of the telescope focal plane, $F\#$ is the focal length of the telescope subsystem, and $\rm{\epsilon = D_{M_2} / D_{M_1} }$ is the ratio between the diameter of the secondary mirror $\rm{D_{M_2}}$ and the primary mirror. Furthermore, $\text{sinc} = sin(u) / u $  is the sinc-function, with
\begin{equation}
   u = \frac{ \pi }{ \lambda F\#} \big[ \boldsymbol{\rm{x}}_{fp} - f \boldsymbol{\Theta} \Big],
   \label{eq:9}
\end{equation}
which characterizes the diffraction of the telescope subsystem over the telescope focal plane, because $f \boldsymbol{\Theta}$ is the plate scale of this subsystem. In Eq. (\ref{eq:8}) and (\ref{eq:9}), the $- e_o \text{exp}(ikz_{fp})$-term is removed, because it is a constant. Then, the angular spectrum of plane waves at the telescope focal plane is obtained via
\begin{equation}
    \boldsymbol{e}_{fp} ( \boldsymbol{\rm{x}}_{fp}, z_{fp} ) = \int \boldsymbol{T}( \boldsymbol{\rm{x}}_{fp} | \boldsymbol{\Theta} ) \boldsymbol{\rm{a}} (\boldsymbol{\Theta}) d \Theta
    \label{eq:10}
\end{equation}
with 
\begin{equation}
    \boldsymbol{T}( \rm{\textbf{x}}_{fp} | \mathrm{\boldsymbol{\Theta}} ) = \frac{1}{\lambda^2 F\#} \Big[ \text{sinc}(u) - \epsilon \text{sinc}( \epsilon u) \Big]
    \label{eq:11}
\end{equation}
being the kernel that describes the mapping between the point source and the corresponding electric field over the telescope focal plane, and $\boldsymbol{a} (\mathrm{\boldsymbol{\Theta}})$ is the vector contain the weights for the angular spectrum of incoming plane waves. 

To preform computer simulations, a numerical adaptation of the presented theory is required. Therefore, we sample both the Field of View (FoV) of the telescope, i.e. the angular spectrum of incoming plane waves, and the electric field over the focal plane. In other words, the vectors $\boldsymbol{a} (\mathrm{\boldsymbol{\Theta}})$ and $\boldsymbol{e}_{fp} ( \rm{\textbf{x}}_{fp}, z_{fp} )$ are sampled at discrete angular and spatial position, and written as a column vectors $\rm{\textbf{a}}$ and $\rm{\textbf{e}}^{(fp)}$, respectively. In this case, we define 
\begin{equation}
    \boldsymbol{\rm{a}} = \big[ a_1, a_2,\dots, a_{M} \big]^{T} ,
    \label{eq:12}
\end{equation}
where the angular sample positions are given by  
\begin{equation}
	\boldsymbol{\rm{\Theta}} = \big[ \Theta_1, \Theta_2, \dots , \Theta_{M} \big]^{T},
	\label{eq:13}
\end{equation}
and $\Theta_{m}$ is the $m$-th element of $\mathrm{\boldsymbol{\Theta}}$ for $m=1,2,...,M$, where $M$ is the total number of angular sample points in the telescope FoV, and $^{T}$ is the transpose. Moreover, at the focal plane we define
\begin{equation}
    \rm{\textbf{e}}^{(fp)} = \big[ e_1, e_2,\dots, e_N \big]^{T} ,
    \label{eq:14}
\end{equation}
where the spatial sample positions are given by  
\begin{equation}
	\rm{\textbf{x}}^{(fp)} = \big[ x_1,x_2, \dots , x_N \big]^{T}.
	\label{eq:15}
\end{equation}
Here, $^{(fp)}$ labels the discretely sampled telescope focal plane, $x_{n}$ is the $n$-th element of $\boldsymbol{\rm{x}}^{(fp)}$ with $n=1,2,...,N$, where $N$ is the total number of spatial sample points at which the field in the telescope focal plane is sampled. In this case, $\boldsymbol{T}( \rm{\textbf{x}}_{fp} | \mathrm{\boldsymbol{\Theta}} )$ becomes the matrix $\rm{\textbf{T}}$, which has dimensions $N \times M$. Matrix $\rm{\textbf{T}}$ now describes the mapping between a discrete angular of the on-sky point source and its corresponding discrete electric field over the telescope focal plane. From now on, a lower case font is used for vectors and upper case font for matrices.

In general, modelling partially coherent optics requires propagating the spatial state of coherence of the incident electric field. The spatial state of coherence of a discrete electric field $\rm{\textbf{e}}$ at the sample points    is given by the (spatial) correlation matrix
\begin{equation}
	\boldsymbol{\rm{E}} = \big< \rm{\textbf{e}} \rm{\textbf{e}}^{\dagger} \big> ,
	\label{eq:16}
\end{equation} 
where $\big< \, \, \big>$ indicates averaging over a representative ensemble, and $^{\dagger}$ is the Hermitian transpose, and total power in the field is proportional to the trace of matrix $\boldsymbol{\rm{E}}$ \cite{Wolf:82,Withington:01,Withington:07}. As a result, we rewrite Eq. (\ref{eq:16})  
\begin{equation}
    \rm{\textbf{E}} = \rm{\textbf{T}} \rm{\textbf{A}} \rm{\textbf{T}}^{\dagger}
    \label{eq:17}
\end{equation}
such that $ \rm{ \textbf{A}} = \big< \boldsymbol{\rm{a}} \boldsymbol{\rm{a}}^{\dagger} \big> $ describes the angular state of coherence in the source plane, i.e. in the far-field of \textit{Herschel}-SPIRE. The matrix $\rm{ \textbf{A}}$ determines both the spatial distribution of the source, and the correlations between the source at the discrete angular positions in the FoV. In this paper, the angular sample step $\Delta \Theta$ is chosen to be smaller than the spatial resolution of the primary mirror, which is the case for most astronomical objects. Therefore, we consider the point sources at the sampled angular positions to be fully spatially incoherent, i.e. matrix $\rm{ \textbf{A}}$ is diagonal matrix \cite{Kano:62,Withington:01}. 

\subsection{\textit{Herschel}-SPIRE instrument}

Equation (\ref{eq:17}) is an important result, because it describes the partially coherent field that is incident on the \textit{Herschel}-SPIRE. We use our HFMF to propagate the correlation matrix of this field, through the instrument, onto the detector plane, where the total power in the field can be measured by the feed-horn bolometers. Here, we provide a short summary of the HFMF, and we adopt the matrix notation of Ref.~\citenum{Lap:22}.

Modal optics replies on notion of modes to map an incident electric field $\rm{\textbf{E}}$ over the input surface, which can be in any state of coherence, to the output surface of the optical system \cite{Withington:01,Withington:04,Withington:07}. The set of modes is characteristic for the optics, and are equivalent to a field propagator used for propagating incident fields through the optics. The HFMF is a numerical version of the functional theory \cite{Withington:01}, which uses the Huygens-Fresnel diffraction integral for obtaining this field propagator, which here is represented by the system transformation matrix, $\boldsymbol{\rm{\overline{H}}}$.

For now, we consider an optical system comprised of $S$ optical surfaces, with $s=1,2,...,S$ labelling the optical surfaces. The system transformation matrix $\boldsymbol{\rm{\overline{{H}}}}$ is obtained by moving from the input surface, through each optical element, to the output surface of the optical system, and is described by:
\begin{equation}
    \boldsymbol{\rm{\overline{H}}} =  \boldsymbol{\Omega}^{S} {\displaystyle \prod_{s=S-1}^{1}} \big(  \boldsymbol{\rm{\overline{G}}}^{(s)} \boldsymbol{\Omega}^{(s)} \big) .
    \label{eq:19}
\end{equation}
In Eq. (\ref{eq:19}), the indexing of the matrices is reversed to respect the sequential ordering of the optical elements in accordance with the adopted matrix notation. The system transformation matrix $\boldsymbol{\rm{\overline{H}}}$ describes how an electric field over the input surface is mapped, through the optics, onto the output surface, at frequency $\nu$. Furthermore, this system transformation matrix is the product of the transmission matrices $\{ \boldsymbol{\Omega}^{(s)} | s=1,2,...,S\}$, and the propagation matrices $\{ \boldsymbol{\rm{\overline{G}}}^{(s)} | s=1,2,...,S-1 \}$. The former describe the phase transforming properties of the optical surfaces, while the latter maps the electric field over an optical surface to the next optical surface. Furthermore, $\bar{}\bar{}\,$ indicates that the system transformation matrix and the propagation matrices are normalized.

Any FTS has two input and two output ports, but for now, we focus on the first FTS port, i.e. the optical coupling from the sky (first input port) to the detector plane (the first output port)\cite{Swinyard:14}. We apply Eq. (\ref{eq:19}) to the 1-D optical model of \textit{Herschel}-SPIRE (see Fig. \ref{fig:1}), and we scan the Michelson FTS. The discrete positions of the movable mirror, i.e. the OPDs, are described by the column vector 
\begin{equation}
    \boldsymbol{\rm{L}} = \Big[L_1, L_2, \dots, L_{N_L}]^{T} ,
    \label{eq:19_2}
\end{equation} 
where $L_j$ is the $j$-th element of $\boldsymbol{\rm{L}}$ for $j = 1,2,\ldots,N_L$, with $N_L$ being the total number of positions in the scan. Equation (\ref{eq:19_2}) implies that the optical configuration of the system changes for each discrete position of the movable mirror. Therefore, for every $j$-th element in $\boldsymbol{\rm{L}}$, Eq. (\ref{eq:19}) has to be evaluated and the corresponding system transformation matrix $\boldsymbol{\rm{\overline{H}}}_{j}$ is calculated. The general operation principles of the FTS can then be represented by a single system transformation matrix
\begin{equation}
    \boldsymbol{\rm{\overline{H}}} = r_1 t_1 \boldsymbol{\rm{\overline{H}}}_{ZPD} + t_2 r_2 \boldsymbol{\rm{\overline{H}}}_{j},
    \label{eq:20}
\end{equation}
where $\boldsymbol{\rm{\overline{H}}}_{ZPD}$ and $\boldsymbol{\rm{\overline{H}}}_{j}$ are the transformation matrices corresponding with ZPD and the $j$-th OPD, or position of the movable mirror, respectively. Furthermore, $r_1$ and $r_2$, and $t_1$ and $t_2$, are the complex field reflection and transmission coefficients associated with the beamsplitter. In general, the complex reflection and transmission coefficients are not necessarily the same, i.e.  $r_1\neq r_2$, and $t_1 \neq t_2$, but here we want to explore the basic operation principles of a partially coherent Michelson interferometer. Therefore, we will assume that the Michelson FTS is operated using an ideal symmetric beamsplitter. In this case, $|r_1| = |r_2| = |r|$ and $|t_1| = |t_2| = |t|$ (an ideal beamsplitter), and $|r|^2=|t|^2 = 1/2$ (a symmetric beamplitter), therefore Eq. (\ref{eq:20}) becomes 
\begin{equation}
    \boldsymbol{\rm{\overline{H}}} = ( \boldsymbol{\rm{\overline{H}}}_{ZPD} + \boldsymbol{\rm{\overline{H}}}_{j} ) / 2
    \label{eq:21}
\end{equation}
which describes the system characteristics of \textit{Herschel}-SPIRE at frequency $\nu_i$ and OPD $L_j$.

Since the spatial dimensions of \textit{Herschel}-SPIRE are finite, its information throughput is limited, and therefore we can use Singular Value Decomposition (SVD) to decompose matrix $\boldsymbol{\rm{\widetilde{H}}}$ into a unique set of orthonormal basis vectors:
\begin{equation}
    \boldsymbol{\rm{\overline{H}}} = \boldsymbol{\rm{\overline{U}}} \hspace{.5mm} \boldsymbol{\rm{\overline{\Sigma}}} \hspace{.5mm} \boldsymbol{\rm{\overline{V}}}^{\dagger}.
    \label{eq:22}
\end{equation}
In Eq. (\ref{eq:22}), $\boldsymbol{\rm{\overline{U}}}$ is an unitary matrix of dimensions $N' \times N'$, with $N'$ being the total number of discrete points over the detector plane, $\boldsymbol{\rm{\overline{V}}}$ is an unitary matrix of dimensions $N \times N$, with $N$ being the total number of discrete points over the telescope focal plane. Matrix $\boldsymbol{\overline{\Sigma}}$ has dimensions $N' \times N$, and only has entries along its diagonal, which are the singular values of $\boldsymbol{\rm{\overline{H}}}$: $\sigma_{j'}$ for $j'= 1, 2, \ldots, \text{min}(N,N')$, where $\sigma_0 \geq \sigma_1 \geq \ldots \geq \sigma_{j'} $. The singular values can attain a maximum value of unity, because matrix $\boldsymbol{\rm{\overline{H}}}$ is power normalized.

From a physical perspective, Eq. (\ref{eq:22}) is interpreted as follows. \textit{Herschel}-SPIRE is described in terms of characteristic orthogonal vectors that span the telescope focal plane (the columns of $\boldsymbol{\rm{\overline{V}}}$), which map one-to-one to a set of characteristic orthogonal vectors spanning the detector plane (the columns of $\boldsymbol{\rm{\overline{U}}}$) with a certain efficiency (the diagonal elements of $\boldsymbol{\rm{\overline{\Sigma}}}$). We define the optical modes of \textit{Herschel}-SPIRE as the collection of individually coherent, but mutually fully incoherent, fields that have nonzero singular values, and we define an optical system as few-mode when for 5-20 optical modes the singular value is larger than an arbitrary chosen value of $10\%$. 

Until now we considered monochromatic light, but the framework can be easily expanded to a polychromatic description. Consider a discrete set of frequencies, which is described by column vector  
\begin{equation}
	\boldsymbol{\rm{\nu}} = [\nu_1, \nu_2, \dots, \nu_{N_i}]^{T} \text{ for } i = 1,2,\dots,N_i,
	\label{eq:1}
\end{equation}
where $\nu_i$ is the $i$-th element of $\boldsymbol{\rm{\nu}}$ that contains $N_i$ frequencies, each separated by a discrete step, $\Delta \nu_i$. In addition, the spectral light originating from the point source will follow some spectral form, and the source spectrum is described by column vector  
\begin{equation}
	\boldsymbol{\rm{b}} = [b_1,b_2,\dots,b_{N_i}]^{T},
	\label{eq:22_2}
\end{equation} 
where $b_i$ is the $i$-th element of $\boldsymbol{\rm{b}}$. Consequently, the spectral correlation matrix of the corresponding electric field over the telescope focal plane is written as 
\begin{equation}
    \boldsymbol{\rm{B}} = b_i \boldsymbol{\rm{E}}.
    \label{eq:22_3}
\end{equation}
The optical modes are the underlying physical entities that propagate the spectral correlation matrix over the telescope focal plane to the detector plane:
\begin{equation}
    \boldsymbol{\rm{B}}^{\prime(x)} =  b_i \boldsymbol{\rm{\overline{H}}} (\boldsymbol{\rm{T}} \boldsymbol{\rm{A}} \boldsymbol{\rm{T}}^{\dagger}) \boldsymbol{\rm{\overline{H}}}^{\dagger},
    \label{eq:23}
\end{equation}
which is obtained by combing Eq. (\ref{eq:17}) and (\ref{eq:22_3}), and using Eq. (\ref{eq:19}) \cite{Withington:01,Withington:07}. Here, $\boldsymbol{\rm{B}}^{\prime(x)}$ is the correlation matrix describing the state of coherence of the spectral field at the detector plane (labelled by $^{\prime}$) along the $\hat{x}$ axis (labelled by $^{(x)}$) at frequency $\nu_i$, where the frequency label $_i$ has been ignored for simplicity. For \textit{Herschel}-SPIRE the number of optical modes, i.e. the number of modes for which $\sigma > 10\%$, is a function of frequency and OPD, i.e. Eq. (\ref{eq:23}) varies with $\nu_i$ and $L_j$. To obtain polychromatic description of \textit{Herschel}-SPIRE, Eq. (\ref{eq:21}) has to be evaluated for each discrete frequency $\nu_i$, and each discrete OPD $L_j$, which will be described in the following section. 


\begin{table}[t]
\caption{Geometrical and optical properties of the optical elements in \textit{Herschel}-SPIRE. These parameters are use to construct the 1-D inline equivalent model shown in Fig. \ref{fig:1}. } 
\label{tab:1}
    \begin{center}  
        \begin{tabular}{|l|l|l|l|l|l|}
        \hline 
\rule[-1ex]{0pt}{1.0ex}  \textbf{Name optical surface} & \textbf{Label} & \textbf{Function} & \textbf{D (mm)} & \textbf{f (mm)} & \textbf{z (mm)} \\ \hline 
\rule[-1ex]{0pt}{1.0ex} Object &  & - & $\infty$ & - & \\ \hline
\rule[-1ex]{0pt}{1.0ex} Telescope primary mirror & M1 & Paraxial lens & 4550 & 1750 & 1588     \\ \hline
\rule[-1ex]{0pt}{1.0ex} Telescope secondary mirror & M2 & Paraxial lens & 308 & -172.599 & 2638 \\ \hline
\rule[-1ex]{0pt}{1.0ex} Telescope focal plane & - & Aperture & 51  & -         & 93    \\ \hline
\rule[-1ex]{0pt}{1.0ex} Common Mirror 3 & CM3 & Paraxial lens & 81  & 195.967             & 211   \\  \hline
\rule[-1ex]{0pt}{1.0ex} Common Mirror 4 / Beam Steering Mirror & BSM  & Aperture& 20 & - & 193 \\ \hline
\rule[-1ex]{0pt}{1.0ex} Common M5 / Image relay & CM5  & Paraxial lens & 111 & 148.789 & 200 \\  \hline 
\rule[-1ex]{0pt}{1.0ex} Spectrometer Mirror 6 / Pick-off mirror & SM6 & Paraxial lens & 11 & 177.011 & 135 \\ \hline
\rule[-1ex]{0pt}{1.0ex} Cold stop & - & Aperture    & 25 & - & 39 \\ \hline
\rule[-1ex]{0pt}{1.0ex} Spectrometer Mirror 7 & SM7 & Aperture   & 52 & - & 58 \\  \hline
\rule[-1ex]{0pt}{1.0ex} Spectrometer Mirror 8 / Relay in &  SM8 & Paraxial lens    & 82 & 109.274  & 175 \\ \hline
\rule[-1ex]{0pt}{1.0ex} Beamsplitter 1 & BS1 & Aperture & 38 & - & 32 \\ \hline
\rule[-1ex]{0pt}{1.0ex} Intermediate image plane 1 & IIP1 & Aperture & 30 & - & 134 \\ \hline
\rule[-1ex]{0pt}{1.0ex} Spectrometer Mirror 9 & SM9 & Paraxial lens    & 66 & 134 & 127\\ \hline
\rule[-1ex]{0pt}{1.0ex} Corner Cube 1 & CC1 & Aperture   & 52 & - & 30\\ \hline
\rule[-1ex]{0pt}{1.0ex} Intermediate pupil image & & Aperture & 30 & - & 30 \\ \hline
\rule[-1ex]{0pt}{1.0ex} Corner Cube 2 & CC2 & Aperture & 52 & -& 127 \\ \hline
\rule[-1ex]{0pt}{1.0ex} Spectrometer Mirror 10 & SM10 & Paraxial lens & 78& 134& 134 \\ \hline
\rule[-1ex]{0pt}{1.0ex} Intermediate image plane 2 & IIP2 & Paraxial lens& 23& -& 40 \\ \hline
\rule[-1ex]{0pt}{1.0ex} Beamsplitter 2 & BS2 & Aperture & 38 & - & 153 \\  \hline
\rule[-1ex]{0pt}{1.0ex} Spectrometer Mirror 11/ Relay out &  SM11 & Paraxial lens & 96 & 96.306 & 70 \\\hline
\rule[-1ex]{0pt}{1.0ex} Detector pupil & - & Aperture & 22 & - & 119 \\ \hline
\rule[-1ex]{0pt}{1.0ex} Spectrometer Mirror 12 & SM12 & Paraxial lens & 22 & 119 & 3   \\  \hline
\rule[-1ex]{0pt}{1.0ex} Detector plane & - & Aperture & 23& - & -  \\ \hline
        \end{tabular}
    \end{center}
\end{table}

\section{Simulation results}
\label{sec:results}

The geometrical and optical parameters of the \textit{Herschel}-SPIRE optical components are presented in Table \ref{tab:1}, where $D$ is the aperture width, $z$ is the distance to the next surface, and $f$ is the focal length.  We used an equidistant sampling of $ \Delta x = c / (2 \nu_{max})$ for all optical surfaces, where where $c$ is the speed of light, and $\nu_{max}= 990\,$GHz is the maximum frequency at which the SLW operated. 

In our simulations, different input and straylight spectra were used to highlight different aspects of \textit{Herschel}-SPIRE. All of these spectra, the specification of which will be described later on, were defined over the SLW frequency range ($447-990$ GHz) and generated following the same procedure \cite{Lap:22}. Furthermore, the spectra were oversampled by a factor of eight with respect to spectral resolution of the FTS $R$ ($\Delta \nu = \Delta \nu_i = 0.25$ GHz) to ensure that i) common, numerical artefacts, such as spectral aliasing, were avoided, ii) the random fluctuations in the fields were narrow in bandwidth $\Delta \nu$ compared to the mean frequency ($\bar{\nu} = 733.5$ GHz) \cite{Wolf:07}, and iii) that the state of coherence of each spectrum was incorporated in accordance with Eq. (\ref{eq:23}). 

In general, the \textit{Herschel} FTS was operated in two modi: single- and double-sided scans. In practise, single-sided scans were used more often, because an higher $R$ can be attained. However, in our simulations we focused on double-sided scans, because it allowed us to demonstrate the applicability of the modal framework. In double-sided scans, the interferogram is measured in both the positive and negative OPD. In our simulations, the discrete OPDs were equidistantly spaced and described by the column vector 
\begin{equation}
    \boldsymbol{\rm{L}} = \Delta z \Big[-\frac{N_L }{2},-\frac{N_L - 1}{2},\ldots,-2,-1,0,1,2,\ldots,\frac{N_L - 1}{2},\frac{N_L}{2} \Big] ^{T} ,
    \label{eq:26}
\end{equation} 
where $\Delta z = 5\mu\,$m is the sampling step for the discrete positions of the movable mirror, and the scan length along each direction was $L = \big| \frac{\Delta z N_L }{2} \big| = 70$mm \cite{Locke:06}, such that $N_L=28001$. 

The simulation results are divided into two parts: the \textit{Herschel}-SPIRE instrument and the end-to-end analysis of \textit{Herschel}-SPIRE. In the first section, we focus on the \textit{Herschel}-SPIRE instrument, i.e. the telescope subsystem and the detector array were excluded, and we analyse its few-mode characteristics, which was typified by the FTS. In the second section, the \textit{Herschel} telescope detector array is added, and the end-to-end partially coherent modelling of \textit{Herschel}-SPIRE is presented. Here, the few-mode frequency-dependent beam of \textit{Herschel}-SPIRE is analysed, and we demonstrate how the HFMF can be used to support the design, verification, and calibration of future spectroscopic FIR missions, in the presence of straylight.

\subsection{Few-mode analysis of the \textit{Herschel}-SPIRE instrument}
\label{sec:results:optics}

\begin{figure} [b]
   \begin{center}
   \begin{tabular}{c} 
   \includegraphics[width=\textwidth]{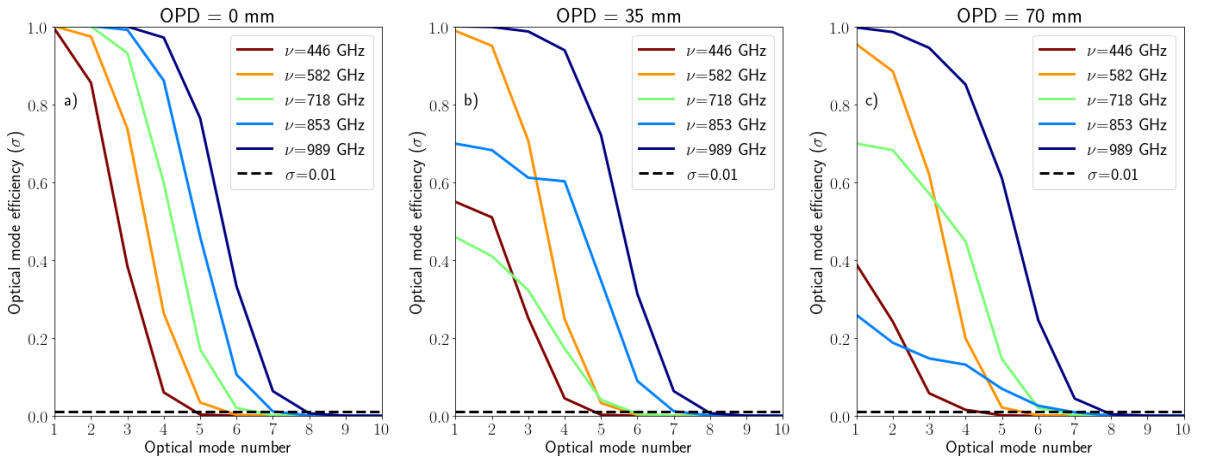}
   \end{tabular}
   \end{center}
   \caption[example] 
   {\label{fig:2} 
Number of optical modes of \textit{Herschel}-SPIRE instrument as function of frequency and OPD.}
\end{figure}

Using the parameters presented in Table \ref{tab:1}, the discrete frequencies from $\boldsymbol{\rm{\nu}}$, the discrete OPDs in $\boldsymbol{\rm{L}}$, and Eq. (\ref{eq:21}), we obtained the normalized system transformation matrix $\boldsymbol{\rm{\overline{H}}}$. Next, we used Eq. (\ref{eq:22}) to determine the optical mode efficiencies for five discrete frequencies and three OPDS. The results are shown in Fig. \ref{fig:2} for the first eight modes.  

In Fig. \ref{fig:2}a), we see that the number of optical modes decrease with frequency. Furthermore, when we compare Fig. \ref{fig:2}a), b) and c), we see that the number of optical modes also varies with OPD. Both of these effects are due to spatial filtering within the two FTS arms. In general, the number of optical modes decrease linearly with frequency and OPD. However, in reality the transition in the number of modes is more complex. For instance, when comparing the curves in Fig. \ref{fig:2} for $\nu_i = 718\,$GHz, we see that the number of optical modes at OPD=$70\,$mm is higher than for OPD=$30\,$mm, but lower for ZPD. These changes in optical mode number are because matrix $\boldsymbol{\rm{\overline{H}}}$ describes the optical coupling and interference between the two arms simultaneously. It is important to model this few-mode FTS behaviour accurately, because later on, when a few-mode detector is coupled to optics, it will determine the partially coherent behaviour of \textit{Herschel}-SPIRE at system level. 

\subsubsection{Straylight}
\label{sec:results:optics:stray}

In ultra-sensitive few-mode FIR instruments, it is important to optimize the instrument optical response to the source, while simultaneously controlling the instruments response to straylight, because straylight can be greatly affected the instrument performance. Before we discuss the straylight analysis of \textit{Herschel}-SPIRE instrument in detail, we need to explain how straylight can be included into the simulations. In this paper, two straylight sources are considered: i) external generated straylight (i.e. thermal background radiation originating from regions surrounding the observed source or the preceding telescope optics), and ii) internal generated straylight (i.e. thermal background radiation from the instrument itself). For now, we will focus on internal straylight, the effects of external straylight will be discussed in Sec. \ref{sec:results:E2E:assess}. Here, we provide a short explanation on how internal instrument straylight was included, see Sec. 2.3.5 of Ref.~\citenum{Lap:22} for more details. 

In the case of internal straylight, we made a number of first-order approximation about the instrument and its mechanical enclosure. First, we assumed that the instrument was enclosed by the same mechanical structure, to ensure that the optical components were supported, and to shield the instrument from external radiation. Second, we assumed that the walls of the enclosure were covered with perfect EM absorber (see Fig. \ref{fig:1}), such that all incident radiation onto the absorber was absorbed, i.e. the enclosure became a perfect black body. Third, and last, we assumed that the perfect black body enclosure and the enclosed optical and mechanical elements were in thermal equilibrium with a physical temperature $\,T_s \sim 5 \,$K, and therefore they emitted incoherent thermal background radiation. This straylight radiation propagated through the optics and was able to reach the detector plane. The straylight correlation matrix $\boldsymbol{\overline{\rm{C}}}'$ that describes the partially coherent straylight field at the detector plane is given by 
\begin{equation}
    \boldsymbol{\overline{\rm{C}}}' = c_i \boldsymbol{\rm{\overline{U}}} (\boldsymbol{\rm{I}}' - \boldsymbol{\rm{\Sigma}}^2 ) \boldsymbol{\rm{\overline{U}}}^{\dagger} ,
    \label{eq:27}
\end{equation}
where $\boldsymbol{\rm{I}}'$ is the identity matrix of dimensions $N'\times N'$, and $c_i$ is the $i$-th element of column vector $\boldsymbol{\rm{c}}$ describing the spectral form the straylight \cite{Lap:22}. The $\boldsymbol{\rm{U}} (\boldsymbol{\rm{I}}' - \boldsymbol{\rm{\Sigma}}^2 ) \boldsymbol{\rm{U}}^{\dagger}$ terms in Eq. (\ref{eq:27}) are the straylight modes for the first FTS port. These modes occur when the efficiencies of the optical modes are smaller than unity, i.e. when there are optical losses in the system.

In the straylight simulations the second FTS port had to be included, because it established a second path for straylight coupling, and this port was used for terminating any common modes that entered the first FTS port, such as internal straylight. In our model, the second FTS port is defined by the optical path from the detector plane, through the optics, to the moveable mirror, and back, where the EM absorber surround the detectors over the detector plane acts both as second input and output port of the FTS. Similar to how Eq. (\ref{eq:21}) was obtained for the first FTS port, the optical coupling of the second FTS port was described by 
\begin{equation}
    \boldsymbol{\rm{\widetilde{H}}} = ( \boldsymbol{\widetilde{\rm{H}}}_{ZPD} + \boldsymbol{\widetilde{\rm{H}}}_{j} ) / 2 ,
    \label{eq:29}
\end{equation}
where $\boldsymbol{\rm{\widetilde{H}}}$ is used the normalized system transformation matrix describing the second port of the FTS. Then, we used the SVD of $\boldsymbol{\rm{\widetilde{H}}}$ to obtain the straylight correlation matrix describing the state of coherence of the straylight field at the detector plane for the second FTS port: 
\begin{equation}
	\boldsymbol{\rm{\widetilde{C}}}' = c_i \boldsymbol{\rm{\widetilde{U}}} (\boldsymbol{\rm{I}}' - \boldsymbol{\rm{\widetilde{\Sigma}}}^2 ) \boldsymbol{\rm{\widetilde{U}}}^{\dagger},
	\label{eq:30}
\end{equation}
which describes the straylight coupling to the detector plane through the second port of the FTS. In Eq. (\ref{eq:30}) the $(\boldsymbol{\rm{I}}' - \boldsymbol{\rm{\widetilde{\Sigma}}}^2 ) \boldsymbol{\rm{\widetilde{U}}}^{\dagger}$-term are the straylight mode associated with the second FTS port. 

At the detector plane the straylight field from the first and second FTS port interfere: 
\begin{equation}
	\boldsymbol{\rm{C}}^{\prime(x)} = \boldsymbol{\rm{\overline{C}}}' + \boldsymbol{\rm{\widetilde{C}}}'.
	\label{eq:30_1}
\end{equation}
where the $^{(x)}$ label is used to indicate that the internal straylight correlation matrix along the $\hat{x}$ is considered. In comparison with the power $\boldsymbol{\rm{\overline{C}}}'$ and $\boldsymbol{\rm{\widetilde{C}}}^{\prime}$ the power in $\boldsymbol{\rm{C}}^{\prime(x)}$ is reduced, due to destructive interference between $\boldsymbol{\rm{\overline{C}}}'$ and $\boldsymbol{\rm{\widetilde{C}}}^{\prime}$. However, a small part of the straylight signal $\boldsymbol{\rm{C}}^{\prime(x)}$ remained, since $\boldsymbol{\rm{\overline{H}}}$ and $\boldsymbol{\rm{\widetilde{H}}}$ are not perfectly symmetric. The power in straylight field $\boldsymbol{\rm{C}}'$ is recorded as function of frequency and OPD, resulting in what we call the straylight interferogram. 

\subsubsection{The optical and straylight response of the \textit{Herschel}-SPIRE instrument}
\label{sec:results:optics:optical_stray_response}

\begin{figure} [b]
   \begin{center}
   \begin{tabular}{c} 
   \includegraphics[width=\textwidth]{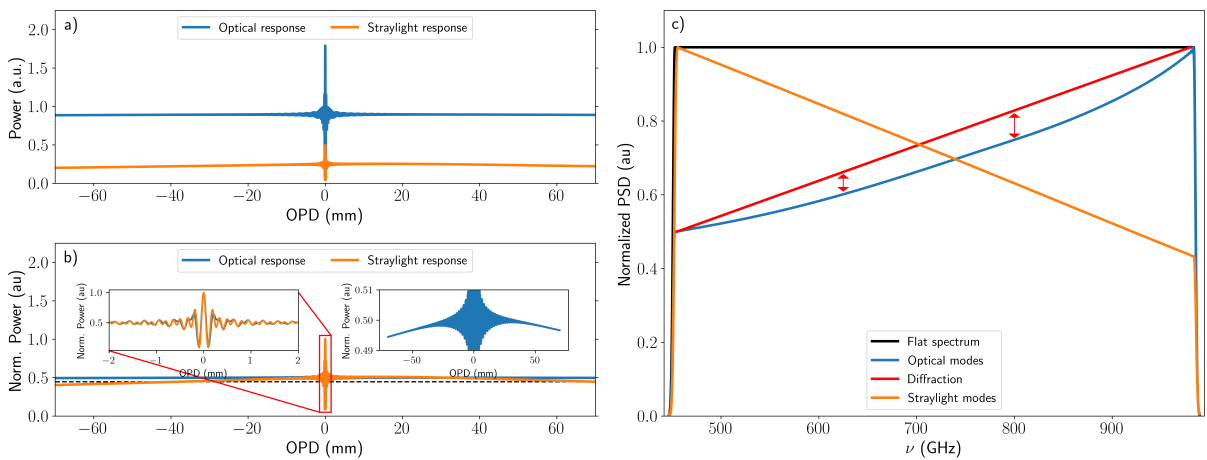}
   \end{tabular}
   \end{center}
   \caption[example] 
   {\label{fig:3} 
Source and straylight interferogram are shown in a), and the normalized inerferograms are shown in b). In b), the upper left inset is a zoom-in on the central peak of the normalized interferogram, and the inset in the upper right a zoom in on the DC-value of the normalized source interferogram is shown, highlighting the asymmetries in the source interferogram. The corresponding spectra are shown in c). Here, the solid black line is the flat source spectrum, the blue indicates the measured source spectrum, the red solid line is shows the spatial filtering of the optics, and the solid orange line the measured straylight spectrum.}
\end{figure}

To study the effects of the optical and straylight modes in the few-mode \textit{Herschel}-SPIRE instrument, we used a flat spectrum for both the source and straylight. In practise this would never occur, because straylight generally follows the PSD a black body \cite{Lap:22}, but it allowed the optical and straylight response of the \text{Herschel}-SPIRE instrument to be compared like-to-like. The flat spectrum had a PSD of unity within the frequency-band, but at the edges of the band it was smoothed with a Gaussian to avoid ringing due to the Gibbs phenomenon \cite{Gibbs:89}. Furthermore, the point source was placed on-axis. The total power in the two fields over the detector plane for frequency $\nu_i$ and OPD $L_j$, i.e. $P_{i,j}^{(ob)}$ and $ P_{i,j}^{(in)}$, respectively, were obtained by taking the trace of correlation matrices $\boldsymbol{\rm{B}}^{\prime(x)}$ and $\boldsymbol{\rm{C}}^{\prime(x)}$, and integrating over spectral bin $\Delta \nu_i$, such that
\begin{equation}
 	    P_{i,j}^{(ob)} = \text{Tr} \Big\{ \boldsymbol{\rm{B}}^{\prime(x)} \Big\} \Delta\nu_i ,
 	    \label{eq:30_2}
\end{equation}
and
\begin{equation}
 	    P_{i,j}^{(in)} = \text{Tr} \Big\{ \boldsymbol{\rm{C}}^{\prime(x)} \Big\} \Delta\nu_i 
 	    \label{eq:30_3}
\end{equation}
In Eq. (\ref{eq:30_2}) and (\ref{eq:30_3}), $^{(ob)}$ labels the astronomical object, i.e. the point source, and the $^{(in)}$ label is used to indicate internal straylight, and $\text{Tr}$ specifies that the trace is taken. Next, the source and straylight interferogram were obtained by integrating $P_{i,j}^{(ob)}$ and $P_{i,j}^{(in)}$ over frequency, and they are shown in Fig. \ref{fig:3}a). Here, we see that the source interferogram (or the optical response) is higher than the straylight interferogram (or straylight response). The reason for this is that the two ports of the FTS interfere destructively, resulting in a lower straylight signal. Figure \ref{fig:3}b) shows the peak-normalized interferograms, such that their overall shape can be compared. From this figure three main observation can be made. First, the optical and straylight around ZPD are different (see the upper left inset in Fig. \ref{fig:3}), because the spatial form the optical and straylight modes are different. Second, the interferograms are asymmetric (see the upper right inset, and the black dashed horizontal line), because the modal content is not symmetric around ZPD. Third, for large negative and positive OPD values the optical response decreases, due to spatial filtering.  

Finally, the normalized measured spectra are shown in Fig. \ref{fig:3}c). From this figure we see that the measured source spectrum suffers from diffraction, because the optical response decrease for lower frequencies (see the red solid line for comparison). However, the losses are not linear (see the red arrows). The reason for this is that the FTS also incorporates the interference between the two FTS arms, which results in addition losses. It is important to understand this modal behaviour in detail, because this broadband few-mode behaviour (the blue line) can be easily misinterpreted as spectral features inherent to the source. On the other hand, the measured straylight spectrum shows the opposite effect, i.e. the straylight response decreases with frequency, because the number of optical modes increase with frequency, which couple to the sky and not to the surrounding enclosure.

These simulations results demonstrated how the framework can be used for describing the partially coherent behaviour of a complex optical systems in detail. We now turn to the end-to-end frequency-dependent partially coherent modelling of \textit{Herschel}-SPIRE.

\subsection{End-to-end partially coherent analysis of \textit{Herschel}-SPIRE}
\label{sec:results:E2E}


This section comprises five parts. In the first part, we describe the 3-D full EM modelling of a single detector in the SWL detector array, and how the 2-D horn aperture mode set over the detector apertures was calculated, which determines the few-mode behaviour of the detector. In the second part, we describe how the 1-D modelling scheme is converted to accommodate 2-D optics, and how 2-D detector model was integrated. In third part, we discuss the spectral calibration scheme. Next, we discussed the 2-D end-to-end partially coherent analysis of \textit{Herschel}-SPIRE. In the fourth part, we present the simulations results of the \textit{Herschel}-SPIRE frequency-dependent beam pattern, and in the fifth, and final, part, we demonstrate how the modal framework can be used in the design, verification, and calibrations strategies of future FIR missions based on three examples.

\subsubsection{Few-mode frequency-dependent detector model}
\label{sec:results:E2E:detector_array}

The SLW detectors were conical feed-horns, which were arranged in a hexagonally close-pack pattern to form the SLW detector array \cite{Makiwa:13,Swinyard:14}. Here, we selected and modelled a single pixel: the central on-axis detector, because all detectors exhibited the same modal behaviour \cite{Makiwa:13}, and it simplified the modelling, but it meant that effects at large off-axis angles were ignored. The SLW conical feed-horn was 3-D and few-mode by design. In our simulations, we characterised the optical response of the detector by a set of orthogonal 2-D reception patterns over its horn aperture. This partially coherent detector reception pattern was described by the frequency-dependent detector correlation matrix $\boldsymbol{\rm{D}}_i$ \cite{Lap:22}. Here,  we describe how matrix $\boldsymbol{\rm{D}}_i$ was obtained. 

\begin{table}[t]
\caption{Detector waveguide modes and their characteristics. } 
\label{tab:2}
    \begin{center}  
        \begin{tabular}{|l|l|l|l|}
        \hline 
        \rule[-1ex]{0pt}{1.0ex}  \textbf{Name} & \textbf{Type} & \textbf{Cut-on $\nu$ (GHz)} & \textbf{Polarizations} \\ \hline 
        \rule[-1ex]{0pt}{1.0ex} $\text{TE}_{11}$ & Transverse Electric & 447 & 2 \\ \hline 
        \rule[-1ex]{0pt}{1.0ex} $\text{TM}_{01}$ & Transverse Magnetic & 584 & 1 \\ \hline 
        \rule[-1ex]{0pt}{1.0ex} $\text{TE}_{21}$ & Transverse Electric & 742 & 2 \\ \hline 
        \rule[-1ex]{0pt}{1.0ex} $\text{TE}_{01}$ & Transverse Electric & 931 & 2 \\ \hline 
        \rule[-1ex]{0pt}{1.0ex} $\text{TM}_{11}$ & Transverse Magnetic & 931 & 1 \\ \hline
        \end{tabular}
    \end{center}
\end{table}

We used the SLW detector parameters listed in Ref.~\citenum{Chattopadhyay:03} to build a 3-D horn model of a single detector in an EM simulator software package \footnote{(2022) CST Microwave Studio. \url{http://www.cst.com} }. In our horn model, we focussed on the waveguide and the conical horn section, and we assumed ideal optical coupling to detector cavity. The 2-D reception patterns over the horn aperture were obtained in the following four steps. In the first step, we used a waveguide port to excite the exit plane within the exit waveguide, and calculate the model content over this 2-D plane. The resulting set of waveguide modes and their characteristics are shown in Table \ref{tab:2}, which are in good agreement with literature values and theoretical calculations \cite{Ramo:12,Makiwa:13}. The cut-on frequency was crucial in the simulations, because this enabled the detector modes in the SLW band with frequency. In the second step, we used a waveguide port to excite the horn aperture plane, and we selected the first 40 horn apertures modes. In the third step, we used the 3-D EM simulator to calculate the scattering coefficients between the set of exit waveguide modes and the horn aperture modes as a function of frequency. In the fourth, and final, step, we obtained the frequency-dependent detector correlation matrix $\boldsymbol{\rm{D}}_i$ via:
\begin{equation}
    \boldsymbol{\rm{D}}_i = \sum_{k'=1}^{40} \sigma_{k',i} \boldsymbol{\rm{d}}_{k'}^{} \boldsymbol{\rm{d}}_{k'}^{\dagger} \text{ for } k'=1,2,\dots,N_m(\nu_i) ,
    \label{eq:31}
\end{equation}
where $\boldsymbol{\rm{d}}_{k'}$ is electric field of the $k'$-th orthogonal detector mode over the horn aperture. The spatial form of the horn aperture modes were calculated using EM theory \cite{Ramo:12}. Furthermore, $\sigma_{k',i}$ is the scattering coefficient between the exit waveguide modes and the horn aperture calculated in step three above. This coefficient controls the modal content over the 2-D horn aperture via the cut-on frequency of the modes, i.e. $\sigma_{k',i}$ is frequency-dependent, as indicated by label $_i$. A subset of the exit waveguide aperture mode were linearly polarized (see Table \ref{tab:2}), there polarization states had to included as individual waveguide mode. However, as stated in Sec. \ref{sec:theory}, we explicitly ignored polarization effects, and therefore were these polarization states added in quadrature. 

\subsubsection{2-D partially coherent modelling framework}
\label{sec:results:E2E:2D}

Until now, 1-D electric fields were considered, but the reception pattern of the detector ewer intrinsically 2-D. In other words, coupling the resulting electric field $\boldsymbol{\rm{B}}'$ directly to the detector is not possible, and an intermediate step is required. We used square separability of the optics to resolve this issues:
\begin{equation}
	\boldsymbol{\rm{B}}^{\prime(ob)} = \boldsymbol{\rm{B}}^{\prime(x)} \times {\boldsymbol{\rm{B}}^{\prime(y)}}^{T},
	\label{eq:32}
\end{equation}
where $\boldsymbol{\rm{B}}^{\prime(x)}$ and $\boldsymbol{\rm{B}}^{\prime(y)}$ are the correlation matrices describing the spectral field over the detector plane along the $\hat{x}$ and $\hat{y}$ axis, respectively. Square separability is valid, because the effective focal length of the \textit{Herschel}-SPIRE is much smaller than the total propagated distance by the electric field within the instrument. Moreover, we verified this assumption by comparing the results from 3-D modal simulations with the results obtained using Eq. (\ref{eq:32}). Next, we assumed that \textit{Herschel}-SPIRE was square symmetric, i.e. $\boldsymbol{\rm{B}}'_{x} = \boldsymbol{\rm{B}}'_{y} $. In our case, this assumption was a good approximation to first order, and Eq. (\ref{eq:32}) was simplified to
\begin{equation}
	\boldsymbol{\rm{B}}^{\prime(ob)} = \boldsymbol{\rm{B}}^{\prime(x)} \times {\boldsymbol{\rm{B}}^{\prime(x)}}^{T}.
	\label{eq:33}
\end{equation} 
As a consequence of Eq. (\ref{eq:32}) and (\ref{eq:33}) is that the FoV is now 2-D instead of 1-D. In other words, the sampled angular positions in the FoV are now described by a $M\times 2$ angular position matrix
\begin{equation}
	\boldsymbol{\rm{\Theta}}_{xy} = [ \boldsymbol{\rm{\Theta}}_x , \boldsymbol{\rm{\Theta}}_y]
	\label{eq:32_1}
\end{equation}
where $\boldsymbol{\rm{\Theta}}_{x}$ and $\boldsymbol{\rm{\Theta}}_{y}$ are the column vectors containing the angular positions along the $\hat{x}$ and $\hat{y}$ axis in the FoV. 

Equations (\ref{eq:32}) and (\ref{eq:32_1}) are important results, because they enable the spatial–spectral performance of \textit{Herschel}-SPIRE, and complex FIR optical systems in general, to be evaluated within a single 2-D theoretical framework. For instance, consider a single point source in the FoV at angular position with index $m$ at frequency $\nu_i$ and OPD $L_j$. The point source creates a secondary field over the telescope primary, and the corresponding correlation matrix of this spectral field is propagated to the detector plane using Eq. (\ref{eq:33}). Here, the resulting field $\boldsymbol{\rm{B}}^{(ob)}_{m}$ is coupled to the central SLW detector, i.e. Eq. (\ref{eq:30_2}) is rewritten into  
\begin{equation}
 	    P_{i,j,m}^{(ob)} = \text{Tr} \Big\{ \boldsymbol{\rm{D}}_i \boldsymbol{\rm{B}}^{\prime(ob)}_{m} \Big\} \Delta\nu_i .
 	\label{eq:34}
\end{equation}
The quantity $P_{i,j,m}^{(ob)}$ is characteristic for the system, and includes the optical coupling between the source and the telescope, the optical properties of the optical elements, such as spatial filtering, and the optical coupling of the resulting spectral field over the detector plane to the detector. In other words, it describes how the flux density of point source  is measured by \textit{Herschel}-SPIRE as a function of angular position ($\boldsymbol{\rm{\Theta}}_{xy}$), frequency ($\boldsymbol{\rm{\nu}}$), and OPD ($\boldsymbol{\rm{L}}$). 

In practise, the point source follows the source spectrum $\boldsymbol{\rm{b}}_{m}$, and the FTS records the interferogram in a single scan, 
\begin{equation}
	\boldsymbol{\rm{p}}_{m}^{(ob)} = \Big[ \sum\limits_{i}^{N_i} P_{i,1,m}^{(ob)}, \sum\limits_{i}^{N_i} P_{i,2,m}^{(ob)}, \ldots, \sum\limits_{i}^{N_i} P_{i,N_j-1,m}^{(ob)}, \sum\limits_{i}^{N_i} P_{i,N_j-1,m}^{(ob)} \Big]^T.
\end{equation}
which the column vector describing the interferogram for source with label $_m$. The interferogram $\boldsymbol{\rm{p}}_{m}^{(ob)}$ is converted into the measured spectrum $\boldsymbol{\rm{b'}}_{m}$ (in $\text{WHz}^{-1}$) via a Fourier Transform. The measured spectrum $\boldsymbol{\rm{b}}^{\prime}_{m}$ is uncalibrated, because $\boldsymbol{\rm{p}}_{m}^{(ob)}$ contains the coupled few-mode behaviour of the optics and the detector, but these effects are removed during calibration.

\subsubsection{Spectrum calibration}
\label{sec:results:E2E:spec_cal}

In \textit{Herschel}-SPIRE, a two-stage calibration pipeline was used to convert the uncalibrated spectrum into a calibrated spectrum \cite{Wu:13,Swinyard:14}. First, an extended source calibration was preformed, and a surface brightness, or “Level-1”, product was obtained. Second, a point source calibration converted the Level-1 product into a flux density, i.e. a “Level-2” product. The former is applicable when the solid angle of the source is much larger than the \textit{Herschel}-SPIRE on-sky beam, while the latter is valid when the solid angle of the source is much smaller than the \textit{Herschel}-SPIRE on-sky beam.

In our simulations, we used the point source calibration only, due to two reasons. First, each interferogram $\boldsymbol{\rm{p}}_{m}^{(ob)}$ was different, because the optical coupling of \textit{Herschel}-SPIRE varied with angular position, frequency, and OPD. Therefore, each $\boldsymbol{\rm{p}}_{m}^{(ob)}$ had to be calibrated individually. Second, the angular spacing between the point sources was chosen to be smaller than the spatial resolution of the telescope. Therefore, an extended source, with an arbitrary spatial and spectral distribution, could be described by a collection of point sources. In the point source calibration, we used a reference source spectrum $\boldsymbol{\rm{b}}^{(ref)}$ of arbitrary spectral form. This spectrum was measured by the FTS, and allowed us to calculate the point source conversion factor, $k_{i,m}$, at discrete frequency $\nu_i$ for each $m-th$ point source in the FoV:
\begin{equation}
	k_{i,m} = \frac{b_{i}^{(ref)}}{b'_{i,m}},
\end{equation}
where $b^{\prime}_{i,m}$ and $k_{i,m}$ are the $i$-th elements of the uncalibrated spectrum $\boldsymbol{\rm{b}}^{\prime}_{m}$ and column vector $\boldsymbol{\rm{k}}^{(ob)}_{m}$, respectively. The latter contains the frequency-dependent point source conversion factors for angular position $m$. Finally, the calibrated measured spectrum $\boldsymbol{\rm{b}}^{\prime\prime}_{m}$ was obtained using
\begin{equation}
	\boldsymbol{\rm{b}}^{\prime\prime}_{m} = \boldsymbol{\rm{k}}^{(ob)}_{m} \odot \boldsymbol{\rm{b}}^{\prime}_{m},
	\label{eq:36}
\end{equation}
where $\odot$ is the element-wise multiplication between the two vectors.

The calibration of the measured spectrum is the last step in our 2-D end-to-end partially coherent modelling framework, and we now turn to applying this scheme to \textit{Herschel}-SPIRE, which consisted out of two parts. In the first part, we use our modal framework analyse the few-mode behaviour of \textit{Herschel}-SPIRE beam pattern. In the second part, we will demonstrate how the modelling technique can be used in the design, verification, and calibration of future FIR instruments.

\subsubsection{The \textit{Herschel}-SPIRE beam pattern}
\label{sec:results:E2E:beam}

\begin{figure} [t]
   \begin{center}
   \begin{tabular}{c} 
   \includegraphics[width=\textwidth]{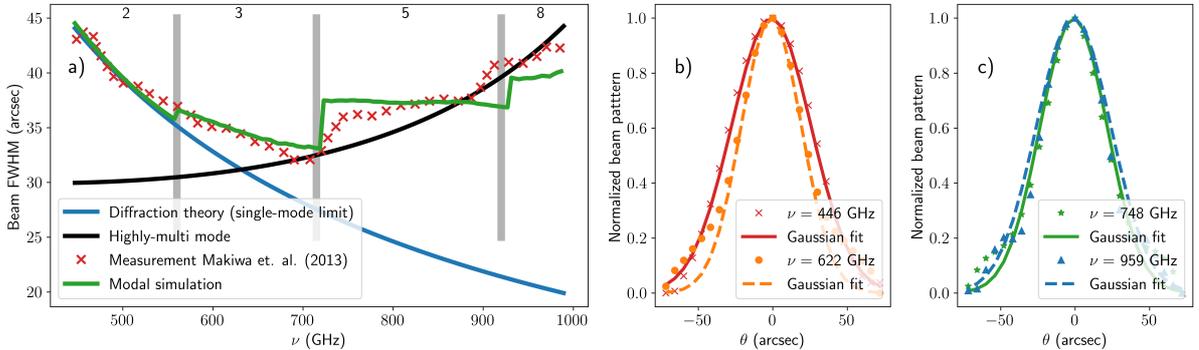}
   \end{tabular}
   \end{center}
   \caption[example] 
   {\label{fig:4} 
Measured (red crosses, from Ref.~\citenum{Makiwa:13}) and simulated (solid green line)\textit{Herschel}-SPIRE beam FWHM as function of frequency (a)). The solid blue line is what is expected from diffraction theory,  i.e. the single-mode limit, and the solid black line shows the trend towards the highly-multi mode limit. The latter increase quadratically, because the number of exit waveguide modes increases quadratically. Furthermore, the vertical grey bars indicate the cut-on frequencies at which the exit waveguide modes are enabled (see Ref.~\citenum{Makiwa:13} and Table \ref{tab:2}), and the number in the top indicate the total number of enabled waveguide modes. b) and c) show the angular cross-cut of the simulated beam for four frequencies, and their Gaussian fit.
    }
\end{figure}

In general, the beam profile of any astronomical instrument must be understood in detail, because it determines how the instrument responds to EM waves originating from an astronomical source. The frequency-dependent beam profile of \textit{Herschel}-SPIRE was determined by obtaining a spectrum at each angular position as a point source is raster-scanned across the FoV, and a Gaussian profile was fitted to determine Full Width at Half Maximum (FWHM) of the beam. The measurement data are shown in Fig. \ref{fig:4}a), which was adopted from Ref.~\citenum{Makiwa:13}\cite{Makiwa:13}. Our aim was to reproduce this results by simulations, i) to confirm that the few-mode beam pattern of \textit{Herschel} was caused by the modal behaviour of the detectors, and ii) to demonstrate that the HFMF can be used to model complex partially coherent optical systems. In our simulations, the unvignetted $2.6'$ instrument FoV was sampled on a $25 \times 25$ rectangular grid. Next, we placed a point source at each grid point position, and we measured an interferogram. The resulting spectra were used to create the simulated frequency-dependent beam profiles, which were fitted with a Gaussian profile to determine their FWHM. The results are shown in Fig. \ref{fig:4}a). 

Figure \ref{fig:4} shows that the measurement and the simulation agree well. At lower frequencies, the system is in the few-mode limit, and the measured, simulated, and the predicated FWHM by diffraction theory agree well. However, when the $\text{TM}_{01}$  mode is enabled, the detector, and consequently the \textit{Herschel}-SPIRE beam, become few-mode. As a result, the FWHM deviates from what is expected from theory, but the simulation result describes the transition well. When we move to higher frequencies, more exit waveguide modes are enabled, and the discrapancy between the measured and theoretical FWHM increases. On the other hand, the modal simulation continues to describe the trend of the measured FWHM well. At higher frequencies, the measured FWHM trends towards the highly multi-mode limit, i.e. the detector reception pattern approach the response function of a patch of EM absorber. This high frequency trend is quadratic, because the number of modes increases quadratically. 

This results confirms that the few-mode behaviour of \textit{Herschel}-SPIRE was indeed determined by the few-mode behaviour of the detectors. Moreover, it shows that the HFMF can describe the trend of the few-mode behaviour, i.e. the transition from the single to highly-multi mode limit, and the frequency at which the exit waveguide mode were enable, accurately, especially when realising that no additional fitting parameters were used. However, a more detailed inspection shows that there are also discrepancies between the measurement and the simulation, two of them are worth noting. First, the measured FWHM is more smooth than the simulation around the cut-on frequencies. The reason for this is that any impedance mismatch between the horn aperture modes and free space, and ohmic losses within the horn and the exit waveguide, are not included in the modal simulation. In principle, these effects should smooth the modal simulation around the cut-on frequencies \cite{Chen:18}, but this is beyond the scope of this work. Another explanation could be optical coupling effects within the back cavity, which were not included in the simulation either. Second, the measured FWHM shows that the modal content of the beam pattern changes around $900$ GHz, while for the simulated this transitions happens later, i.e around $930$ GHz. A possible explanation is that actual geometrical properties of the detector deviated from the design values, causing cut-on the frequency of higher order exit waveguigde modes to changes. Higher order exit waveguide modes are more sensitive to the detector geometry than lower order modes, because their operation wavelength is lower. 

Finally, Fig. \ref{fig:4}b) and c) show that the Gaussian beam fit to the cross-cut of the simulated beam patterns agree well. We can see that the the fit is better for lower than for higher frequencies. At lower frequencies, \textit{Herschel}-SPIRE is operating in the single-mode limit, an therefore the beam is well described by a Gaussian beam \cite{Goldsmith:98}. However, for higher frequencies, the system is few-mode, and a Gaussian fit is no longer appropriate.

\subsubsection{Assessing design, verification, and calibration strategies for future FIR instruments}
\label{sec:results:E2E:assess}

In Sec. \ref{sec:results:optics:stray}, it was mentioned that straylight can greatly affect the performance of ultra-sensitive few-mode FIR instruments. In this paper two straylight sources were considered, i.e. external and internal generated straylight. Until now, internal straylight was included on the instrument level. In this section, we address external and internal straylight on a system level, and we will demonstrate how the modal framework could be used in the design, verification, and calibrations strategies of current and future FIR missions based on three simulation cases.

Before we can discuss these three simulation cases, we need to explain how internal and external straylight were coupled to the detector. In principle, Eq. (\ref{eq:32}) - (\ref{eq:34}) also hold for external and straylight, and by including the contributions from external and internal straylight, Eq. (\ref{eq:34}) is rewitten into 
\begin{equation}
 	    P^{(tot)}_{i,j} = \text{Tr} \Big\{ \boldsymbol{\rm{D}}_i \boldsymbol{\rm{B}}^{\prime(ob)} +  \boldsymbol{\rm{D}}_i \boldsymbol{\rm{B}}^{\prime(ex)} +  \boldsymbol{\rm{D}}_i \boldsymbol{\rm{C}}^{\prime(in)} \Big\} \Delta\nu_i.
 	\label{eq:35}
\end{equation}
Here, label $^{(tot)}$ is used to indicate that the detector measures the total power $P$ for single frequency $\nu_i$ and OPD $L_j$, i.e. the optical coupling to the target, and internal and external straylight, simultaneously. Furthermore, $\boldsymbol{\rm{\overline{B}}}^{\prime(ob)}$ is the target object of the observation, and the $_{m}$ label has been dropped, because an astronomical object can be described as a collection of point sources. Last, $\boldsymbol{\rm{B}}^{\prime(ex)}$ and $\boldsymbol{\rm{C}}^{\prime(in)}$ are the correlation matrices describing the 2-D spectral field over the detector plane due to external (labelled by $^{(ex)}$ and internal (labelled by $^{(in)}$) straylight, respectively:
\begin{equation}
	\boldsymbol{\rm{\overline{B}}}^{\prime(ex)} = \boldsymbol{\rm{\overline{B}}}^{\prime(ex)}_{x} \times {\boldsymbol{\rm{\overline{B}}}^{(ex)}_{x}}^{T}.
\end{equation}
and 
\begin{equation}
	\boldsymbol{\rm{\overline{C}}}^{\prime(in)} = \boldsymbol{\rm{\overline{C}}}^{\prime(in)}_{x} \times {\boldsymbol{\rm{\overline{C}}}^{\prime(in)}_{x}}^{T}.
\end{equation}

\begin{figure} [b!]
   \begin{center}
   \begin{tabular}{c} 
   \includegraphics[width=.75\textwidth]{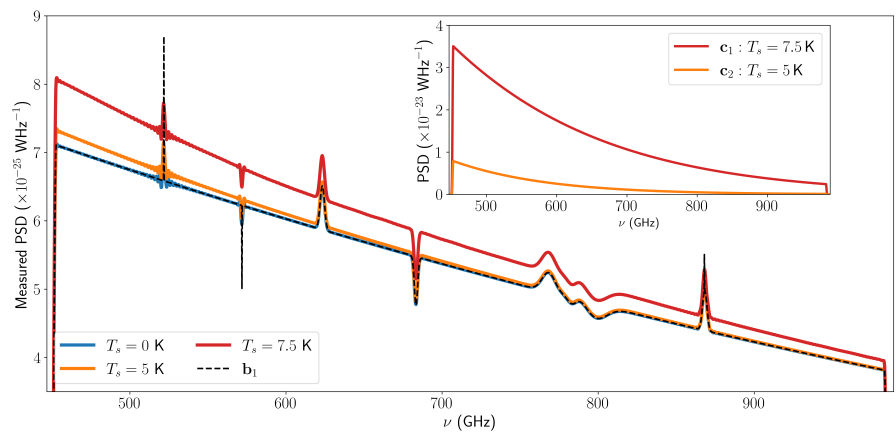}
   \end{tabular}
   \end{center}
   \caption[example] 
   {\label{fig:7} 
	Simulation results for internal straylight: the measurement of a source spectrum ($\boldsymbol{\rm{b}}_2^{(ob)}$, the black dashed line) in different straylight environments. The measured PSD are shown for: i) $T_s = 0\,$K (blue solid line, no straylight), ii) $T_s = 5\,$K (solid orange line), and iii) $T_s = 7.5 \,$K (solid red line). The straylight spectra, i.e. $\boldsymbol{\rm{c}}_1^{(in)}$ and $\boldsymbol{\rm{c}}_2^{(in)}$ are shown in the inset.	
}
\end{figure}

In principle, the calibration procedure described in Sec. \ref{sec:results:E2E:spec_cal} can be used to calibrate the measured spectrum  that is measured through Eq. (\ref{eq:35}). Here, the measured spectrum will vary depending on the contributions for the source, and internal and external straylight, i.e. the $\boldsymbol{\rm{B}}^{\prime(ob)}$, $\boldsymbol{\rm{B}}^{\prime(ex)}$, and $\boldsymbol{\rm{C}}^{\prime(in)}$-term in Eq. (\ref{eq:35}), respectively. We used three representative simulation cases to study the system level effects of straylight: a single case for external straylight, and two cases for internally generated straylight. Furthermore, we used two example source spectra ($\boldsymbol{\rm{b}}_1^{(ob)}$ and $\boldsymbol{\rm{b}}_2^{(ob)}$, see Fig. \ref{fig:5}b)), one external straylight spectrum ($\boldsymbol{\rm{b}}_3^{(ex)}$, see Fig. \ref{fig:6}), and two internal straylight spectra ($\boldsymbol{\rm{c}}_1^{(in)}$ and $\boldsymbol{\rm{c}}_2^{(in)}$, see the insets in Fig. \ref{fig:7}). Here, for clarity, we used $\boldsymbol{\rm{b}}$ to indicate that the spectral light enters the instrument at the telescope focal plane, i.e. secondary spectral source fields and external straylight, while $\boldsymbol{\rm{c}}$ is used to indicate spectral light from within the instrument itself, i.e. internal straylight, and their spectral forms were chosen such that they highlighted different aspects. Last, the source and straylight spectra were generated using the procedure described Ref.~\citenum{Lap:22}, with one difference: the source spectra the Planck's function was used to create the thermal background continuum. 

Source spectrum $\boldsymbol{\rm{b}}_1^{(ob)}$ was a flat spectrum, similar to the one used in Sec. \ref{sec:results:optics:optical_stray_response}, but flux density level of $\boldsymbol{\rm{b}}_1^{(ob)}$ was defined such that it matched the flux density level of $\boldsymbol{\rm{b}}_2^{(ob)}$ at the high-end of the frequency band. The second source spectrum ($\boldsymbol{\rm{b}}_2^{(ob)}$) was a representative astronomical spectrum (see Fig. \ref{fig:5}b)), which consisted of seven representative spectral elements: a thermal background continuum (i) and the red dashed line); narrow (unresolved) emission and absorption lines (ii) and ii)); broad (resolved) emission and absorption features (iv) and v)); and composite features (vi) and vii)). The straylight spectra were controlled by the straylight temperature $T_s$. In the simulations we used $T_s = 7.5, 10 \,$K for internal straylight from the instrument, and we used $T_s=84\,$K for thermal background emission from the telescope \cite{Fulton:14}. The spectra and their spectral features are listed in Table \ref{tab:3}.

\begin{table}[t]
    \centering
    \caption{ Source and straylight spectra specifications. The narrow (unresolved) and broad (resolved) spectral features are indicate by $\delta$ and $\mathcal{N}$, which are a $\delta$-function and a Gaussian distribution, respectively. Moreover, $\boldsymbol{\rm{b}}$ indicates that the spectral light enters the instrument at the telescope focal plane, while $\boldsymbol{\rm{c}}$ is used to indicate spectral light from within the instrument itself, i.e. internal straylight. Finally, $^{(ob)}$ labels the source spectra, $^{(ex)}$ is used to label the internal straylight spectrum, and $^{(in)}$ is used to indicate the internal straylight spectra.}
    \begin{tabular}{|l|l|l|l|l|l|l|} 
        \hline
\textbf{Spectrum} & \textbf{Name} & \textbf{Feature} & \textbf{Type} & $\boldsymbol{\nu}$\textbf{(GHz)}  & $\boldsymbol{\sigma(\nu)}$ & $\boldsymbol{T}$\textbf{(K)} \\ \hline 
$\boldsymbol{\rm{b}}_1^{(ob)}$ & Flat spectrum &  Con. & Emission      & $[\nu_{min},\ldots,\nu_{max}]$ & - & 25 \\ \hline
$\boldsymbol{\rm{b}}_2^{(ob)}$     & Background & Con.     & Emission      & $[\nu_{min},\ldots,\nu_{max}]$ & -& 25\\
& Unresolved line & $\mathcal{N}(\nu,\sigma)$  & Emission    & $521.8$       & $\delta$-function & 30 \\
& Unresolved line & $\delta(\nu)$       & Absorption    & $57.81$       & $\delta$-function    & 22  \\
& Broad line & $\mathcal{N}(\nu,\sigma)$       & Emission      & $623.3$       & $15$    & 26.6 \\
& Broad lne & $\mathcal{N}(\nu,\sigma)$       & Absorption      & $683.3$       & $15$    & 23.2 \\
& Composite & $\mathcal{N}(\nu,\sigma)$ & Absorption      & $[783.4,798.3]$       & $[4,10]$    & $[24.8,24.5]$ \\
&  & $\mathcal{N}(\nu,\sigma)$ & Emission      & $[768.3,788.3]$       & $[6,8]$    & $[25.75,25.1] $ \\
& Composite & $\mathcal{N}(\nu,\sigma)$       & Emission    & $868$       & 2 & 26.5  \\
& & $\delta(\nu)$       & Emission    & $184.9$       & $\delta$-function    & 28  \\ \hline
$\boldsymbol{\rm{b}}_3^{(ex)}$ & Ext. straylight    & Con.	    & Emission      & $[\nu_{min},\ldots,\nu_{max}]$ & - & 84 \\ \hline 
$\boldsymbol{\rm{c}}_1^{(in)}$ & Int. straylight    & Con.	    & Emission      & $[\nu_{min},\ldots,\nu_{max}]$ & - & 5 \\ \hline
$\boldsymbol{\rm{c}}_2^{(in)}$ & Int. straylight    & Con.	    & Emission      & $[\nu_{min},\ldots,\nu_{max}]$ & - & 7.5 \\ \hline
    \end{tabular}
    \newline
    \label{tab:3}
\end{table}

\subsubsection*{External straylight}
In first two simulation cases, we focussed on external straylight, i.e. the $\boldsymbol{\rm{C}}^{\prime(in)}$-term in Eq. (\ref{eq:35}) was ignored. As mentioned, the $\boldsymbol{\rm{B}}^{\prime(in)}$ term in Eq. (\ref{eq:35}) describes external straylight at the detector plane, which opposed to internal straylight, enters the instrument at the telescope focal plane. In other words, Eq. (\ref{eq:23}) must be used to propagate the correlation matrix describing the straylight fields over the telescope focal plane, $\boldsymbol{\rm{B}}^{(ex)}$, to the detector plane. Then, Eq. (\ref{eq:35}) can be applied to measured the interferogram, and the corresponding spectrum can be obtained. 

\begin{figure} [t]
   \begin{center}
   \begin{tabular}{c} 
   \includegraphics[width=.75\textwidth]{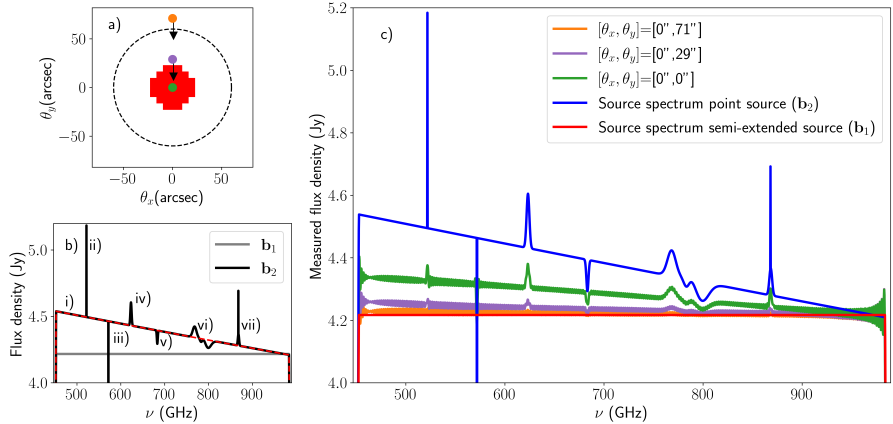}
   \end{tabular}
   \end{center}
   \caption[example] 
   {\label{fig:5} 
    Simulation results for first external straylight case: how the measurement of a semi-extended source spectrum ($\boldsymbol{\rm{b}}_1^{(ob)}$, see b)) was affected by a bright point source ($\boldsymbol{\rm{b}}_2^{(ob)}$, on average $\boldsymbol{\rm{b}}_2^{(ob)} \sim 100 \times \boldsymbol{\rm{b}}_1^{(ob)})$) was scanned into the FoV (see c)). The semi-extended source as defined as a the incoherent superposition of point sources, and placed in the center of the FoV (see b)). Furthermore, the semi-extended source spectrum $\boldsymbol{\rm{b}}_1^{(ob)}$ is a flat spectrum (see b)), and scaled such that the flux densities of $\boldsymbol{\rm{b}}_1^{(ob)}$ and $\boldsymbol{\rm{b}}_2^{(ob)}$ at high frequencies would match. For the point source spectrum $\boldsymbol{\rm{b}}_2^{(ob)}$ was used, which was a representative astronomical spectrum (see Fig. \ref{fig:4}a)) consisting of seven representative spectral elements: a thermal background continuum (i) and the dashed red line); narrow (unresolved) emission and absorption lines (ii)) and (ii)); broad (resolved) emission and absorption features (iv)) and v); and composite features (vi)) and vii)).
}
\end{figure}

In the first external straylight case, we analysed how the measurement of a semi-extended source spectrum was affected when a bright point source was scanned into the FoV (see Fig. \ref{fig:5}a)). Here, we used $\boldsymbol{\rm{b}}_1^{(ob)}$ as the semi-extended source spectrum and $\boldsymbol{\rm{b}}_2^{(ob)}$ as the source spectrum for the point source. Furthermore, on average $\boldsymbol{\rm{b}}_2^{(ob)} \sim 100 \times \boldsymbol{\rm{b}}_1^{(ob)}$, such that the point source was bright compared to the target source (the semi-extended source). The semi-extended source was defined as the incoherent superposition of point sources, each following the same spectral form. In other words, the spatial distribution of point sources on the sky defined the spatial distribution of the semi-extended source, while each of these point sources spectrum followed the spectral form of $\boldsymbol{\rm{b}}_1^{(ob)}$. As in a real observation, the target source was positioned in the center of the FoV (see Fig. \ref{fig:5}b)).

The results of this simulation are show in Fig. \ref{fig:5}, we see that when the bright point source is at the edge of the FoV the spectrometer measures the source spectrum from the semi-extended source. However, when the point source is scanned towards the FoV center, the measured spectrum is a superposition of the semi-extended source spectrum ($\boldsymbol{\rm{b}}_1^{(ob)}$) and the point source spectrum ($\boldsymbol{\rm{b}}_2^{(ob)}$), and we see that the spectral features of $\boldsymbol{\rm{b}}_2^{(ob)}$ start to appear. However, these spectral features $\boldsymbol{\rm{b}}_2^{(ob)}$ are flattened off, due to the superposition of the spectra. Moreover, we can see that the spectrometer has affected the spectrum. For instance, the is broadband ripple introduced, due to the Gibbs phenomenon \cite{Gibbs:89}, and the FTS was unable to resolve the (unresolved) narrow lines due to its limiting resolving power.

In this first external straylight case, we focussed on external straylight from a bright source in the FoV, but many other related system performance analyses are possible. Here, we identified two examples. The first example is observing a (bright) on-axis point source in the presence of thermal background radiation surrounding the target source, i.e. reversing the observation discussed above. This example is frequently encountered, because astronomical sources, i.e. point sources, are generally observed against a (thermal) background. As a result, the point source spectrum is flattened off (comparing the blue and green line in Fig. \ref{fig:5}a)), due to thermal background radiation. The second example, is telescope pointing, which in \textit{Herschel}-SPIRE was one of the largest source of uncertainty \cite{Swinyard:14,Hopwood:15}. In this end-to-end analysis, one evaluates how telescope pointing offsets can affect the measured spectrum, which due to few-mode behaviour of the beam can be complex, especially when the beam is small.

\begin{figure} [t]
   \begin{center}
   \begin{tabular}{c} 
   \includegraphics[width=.6\textwidth]{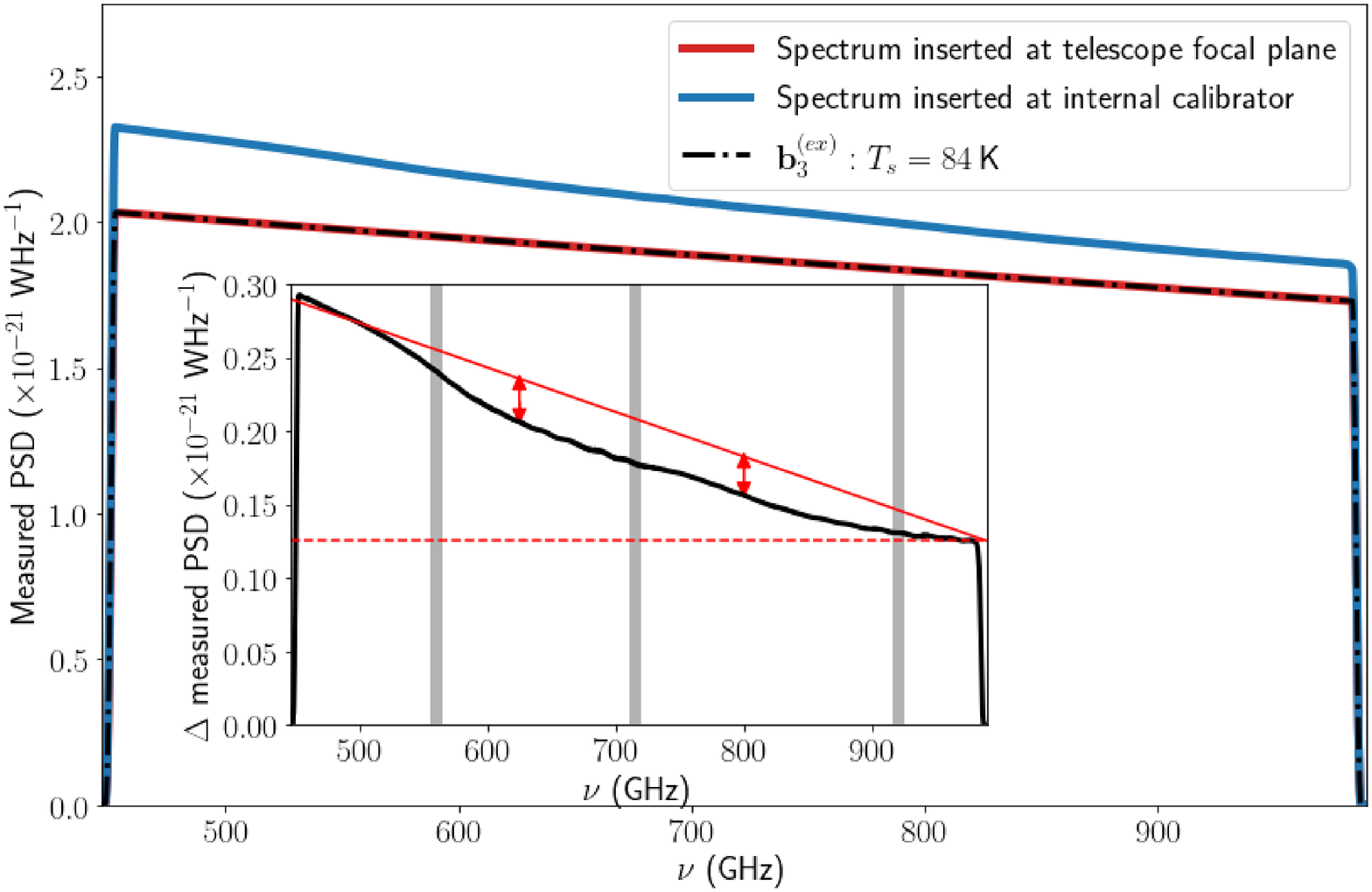}
   \end{tabular}
   \end{center}
   \caption[example] 
   {\label{fig:6} 
    Simulation results for second external straylight case: how to best calibrate thermal background emission from the telescope subsystem. The measured PSDs for the telescope straylight inserted at the telescope focal plane (solid red line) and at the internal calibrator (solid blue line), respectively, for a given external straylight spectrum ($\boldsymbol{\rm{b}}_3$, the black dashed-dotted line. The inset shows the difference between the PSDs (black solid line), where the horizontal dashed line shows the vertical offset, and the solid red line is used to highlight the difference due to spatial filtering.
}
\end{figure}

In the second external straylight case, we studied how to best calibrate for (strong) thermal background radiation emitted by the optical components in front of the instrument, i.e. the telescope subsystem. We used the \textit{Herschel} telescope as an example, because the total signal in the majority of the FTS observations was dominated by the thermal emission from the primary and secondary mirror, and proper calibration was required to remove it \cite{Swinyard:14}. The \textit{Herschel} telescope subsystem was well described by a composite black body emissivity model, i.e. the telescope emission model, with an average physical temperature of $\sim 84\,$K \cite{Fulton:14}. In an earlier design of \textit{Herschel}-SPIRE an internal calibrator was implemented (position at the same position as SM7) to ensure that the thermal radiation from the telescope could be calibrated. However, in a later design iteration the internal calibrator was removed, and the telescope emission model was used for the telescope subsystem calibration. From a calibration perspective, in principle, both approaches can be used, we used the HFMF to evaluate and compare both calibration strategies. Here, for simplicity, we assumed the telescope emission model to be a single black body with $T_s = 84\,$K, and we simulated the thermal background emission from the telescope subsystem using straylight spectrum $\boldsymbol{\rm{b}}_3^{(ex)}$ (see Table \ref{tab:3}). The simulated thermal background radiation from the telescope subsystem, i.e. it was fully incoherent, and therefore its correlation matrix $\boldsymbol{\rm{B}}_3^{(ex)}$ was diagonal. This correlation matrix was inserted at two positions within the system: i) at the telescope focal plane, ii) at the internal calibrator, and using Eq. (\ref{eq:35}) the corresponding interferograms were recorded. These interferogram were converted in to spectra via the Fourier Transform. The telescope focal plane measurement was used a reference, i.e. this spectrum was used to obtain the conversion factor $\boldsymbol{\rm{k}}^{(ex)}$ (see Eq. (\ref{eq:36})), to calibrated the two measured spectra. The results for the second internal straylight case are shown in Fig. \ref{fig:6}. 

From Fig. \ref{fig:6} we can see that the measured PSDs are different, which is shown in more detail in the inset. Here, three observations can be made: i) the there is vertical offset between the spectra (the dashed red line horizontal line), ii) to first order the PSD difference (the inset of Fig. \ref{fig:6}) decreases linear with frequency, iii) the observed trend in ii) changes with frequency (the red arrows in the inset of Fig. \ref{fig:6}). The first and second observation can be explained by the difference in spatial filtering, such as truncation and diffraction, along the two optical paths: i) from the telescope focal plane to the detector plane, and ii) from the internal calibrator to the detector plane. The impact of this effect increases for lower frequencies, because there is more diffraction at these frequencies, which is confirmed by Fig. \ref{fig:6}. The third observation is due to the few-mode behaviour of \textit{Herschel}-SPIRE, similar to what was seen in Fig. \ref{fig:3}c) and \ref{fig:4}. The fully incoherent spectral field is filtered by the few-mode optics, and over the detector plane this field is no longer fully coherent, but has transformed into a partially coherent field \cite{Lap:22}. The field state of coherence of the field over the detector plane varies with optical path, due to a different spatial filtering along the optical path. Furthermore, we see that this behaviour is frequency-dependent and related to the cut-on frequencies of the exit waveguide modes (the vertical grey lines are the same cut-on frequencies as in Fig. \ref{fig:4}). As a result, the optical coupling of $\boldsymbol{\rm{B}}_3^{(ex)}$ to the detector is optical path dependent, i.e. the internal calibration source can not be used to calibrate for $\boldsymbol{\rm{B}}_3^{(ex)}$. In this case, the proper calibration strategy is to use the telescope emission model, as was done for \textit{Herschel}-SPIRE. 

\subsubsection*{Internal straylight}
In last simulation case we focussed on internal straylight, i.e. the $\boldsymbol{\rm{B}}^{\prime(ex)}$-term in Eq. (\ref{eq:35}) was ignored. In this case, we simulated how the measurement of a representative astronomical spectrum ($\boldsymbol{\rm{b}}_2^{(ob)}$ ), originating from an on-axis point source, was affected by different internal straylight environments. Here, we use a straylight temperature $T_s=5, 7.5\,$K, i.e. straylight spectrum $\boldsymbol{\rm{c}}_1^{(ex)}$ and $\boldsymbol{\rm{c}}_2^{(ex)}$, to simulate two straylight environments, which were compared to a measurement in the absence of straylight ($T_s=0\,$K). 

Figure \ref{fig:7} shows that measured spectrum in the absence of straylight (the blue solid line) agrees well with the source spectrum $\boldsymbol{\rm{b}}_2$ (the black dashed line). However, the FTS was unable to resolve the (unresolved) narrow lines due to its limiting spectral resolving power and around these features the Gibbs phenomenon is observed. More spectral power is measured at higher frequencies when straylight is introduced (see the orange and red line for $T_s=5\,$K and $T_s=7.5\,$K, respectively), because $\boldsymbol{\rm{c}}_1^{(in)}$ and $\boldsymbol{\rm{c}}_2^{(in)}$ increase towards lower frequencies. In addition, for $T_s=7.5\,$K we observe that the spectrum is also shifted vertically, i.e. more spectral power is measured of the entire band. A similar measurement was carried out during the in-flight calibration of \textit{Herschel}-SPIRE, where an uncertainty in the continuum level was expected, due to a residual in the background \cite{Swinyard:14}. Although a one-to-one comparison between our simulations and the \textit{Herschel}-SPIRE is not possible, it is interesting to see that the overall low-frequency slope of measured spectrum in Fig. \ref{fig:7} is in good agreement with what was observed in \textit{Herschel}-SPIRE to first order. 

\section{Conclusions}

In this paper, we used our modal framework: i) to analyse the behaviour of a few-mode FTS, and ii) to demonstrate how the technique can be used as an end-to-end instrument simulator for future few-mode FIR spectrometers in the presence of straylight. \textit{Herschel}-SPIRE was used as a case study, because it optical design and few-mode behaviour are representative for current and future few-mode FIR spectrometers. 

We studied the few-mode behaviour of the \textit{Herschel}-SPIRE instrument by excluding the detector, which was typified by the modal behaviour of the FTS, as a function of frequency and OPD. Furthermore, we analysed the optical and straylight response of the instrument. The interferograms and measured spectra were analysed, and their few-mode behaviour was explained using the optical and straylight modes, providing insight into the optical coupling mechanism within a complex optical system, such as \textit{Herschel}-SPIRE. 

Next, the \textit{Herschel} telescope and the few-mode detector was added, and the 2-D end-to-end frequency-dependent partially coherent modelling of \textit{Herschel}-SPIRE was performed, which consisted of two parts. In the first part, the few-mode beam FWHM was simulated and compared with the measured frequency-dependent \textit{Herschel}-SPIRE beam, which was few-mode. Our simulation showed that i) the few-mode behaviour of \textit{Herschel}-SPIRE was indeed determined by the few-mode behaviour of the detectors, and ii) that the HFMF simulation agreed well with the measurement. This an important result, because based on this we conclude that the modal framework can be used for the 2-D end-to-end frequency-dependent partially coherent modelling of a complex few-mode optical systems. In the second part, we demonstrated how external and internal straylight is incorporated in the framework, and how the technique can be used during the design, verification, and calibration of future FIR space-based missions. Here, we simulated the effects of external straylight surrounding the target object, external thermal background radiation emitted from the telescope subsystem, and internal straylight generated by the instrument itself. 

Based on these results, we conclude that the modal framework is well suited for the partially coherent modelling of a few-mode FTS, and the end-to-end modelling of complex few-mode FIR systems, such as \textit{Herschel}-SPIRE. In principle, the technique can used to model any few-mode optical system, because it relies on the optical modes of the system, which naturally account for the optical characteristics of the system. This adaptability makes the technique useful for the design, verification, and calibration of future few-mode FIR spectroscopic space missions, which was demonstrate in this paper by the level of detail and breadth of the simulations, both on subsystem an system level. 

We see a great wealth of applications for the HFMF in optics in general, but here we limit ourselves to the applications for the next generation of ultra-sensitive FIR space-based missions. On an instrument level, we can identify the two main application of the method. First, the framework can help inform on the few-mode behaviour of the complex few-mode spectrometers, such as an FTS. This was exemplified by the broadband few-mode behaviour of the \textit{Herschel}-SPIRE, which could ave been misinterpreted as spectral characteristics of the source. The second important application of the modal framework is as an design tool for tailoring the spectral-spatial instrument response, by designing the optical modes of the system. For instance, for assessing how system, instrument, and optical component level design decisions (such as spatial and spectral resolution, or the geometrical and optical properties of different components) drive the mission science goals. On a system level, example applications are i) evaluating the impact of telescope pointing, ii) assessing different calibration strategy for removing thermal radiation emitted by the optical components along the optical train, and internal straylight. The latter is particularly interesting, because the framework offers a clear direction for studying the coupling mechanisms and impact of straylight in a system level performance context. For instance, the straylight performance of each optical subsystem can be evaluated individually, while including the thermal background emission from optically thick optical elements is straightforward. 

In a future paper, we will use the modal framework for the experimental characteristics of optical systems. For instance, we will report on how the modal framework was used for determining determining the system transformation matrix $\boldsymbol{\rm{\overline{H}}}$ for a pair of limiting slits at $104$ GHz, which showed good agreement. Furthermore, we will the modal framework for the modal characterization of a low-resolution grating spectrometer ($R\sim300$), and a medium few-mode FTS ($R\sim1000$), including straylight. Finally, we will extend the modal framework to include polarization to enable partially coherent modeling of the next-generation of high-resolution spectrometers ($R\sim10000$), such a post-dispersed polarizing FTS and Fabry-Pérot interferometers.

\bibliography{references} 

\begin{thebibliography}{10}

\bibitem{Lap:22}
Lap, B. N.~R., Withington, S., Jellema, W., and Naylor, D.~A., ``Modeling the
  partially coherent behavior of few-mode far-infrared grating spectrometers,''
  {\em J. Opt. Soc. Am. A}~{\bf 39},  1218--1235 (Jul 2022).

\bibitem{Duncan:19}
Farrah, D., Smith, K.~E., Ardila, D., Bradford, C.~M., DiPirro, M.~J.,
  Ferkinhoff, C., Glenn, J., Goldsmith, P.~F., Leisawitz, D.~T., Nikola, T.,
  Rangwala, N., Rinehart, S.~A., Staguhn, J.~G., Zemcov, M., Zmuidzinas, J.,
  Bartlett, J., Carey, S.~J., Fischer, W.~J., Kamenetzky, J.~R., Kartaltepe,
  J., Lacy, M.~D., Lis, D.~C., Locke, L.~S., Lopez-Rodriguez, E., MacGregor,
  M., Mills, E., Moseley, S.~H., Murphy, E.~J., Rhodes, A., Richter, M.~J.,
  Rigopoulou, D., Sanders, D.~B., Sankrit, R., Savini, G., Smith, J.-D., and
  Stierwalt, S., ``{Review: far-infrared instrumentation and technological
  development for the next decade},'' {\em Journal of Astronomical Telescopes,
  Instruments, and Systems}~{\bf 5}(2),  1 -- 34 (2019).

\bibitem{kamp:21}
Kamp, I., Honda, M., Nomura, H., Audard, M., Fedele, D., Waters, L. B. F.~M.,
  Aikawa, Y., Banzatti, A., Bowey, J., Bradford, M., and et~al., ``The
  formation of planetary systems with spica,'' {\em Publications of the
  Astronomical Society of Australia}~{\bf 38},  e055 (2021).

\bibitem{Roelfsema:2018}
Roelfsema, P.~R., Shibai, H., Armus, L., Arrazola, D., Audard, M., Audley,
  M.~D., Bradford, C., Charles, I., Dieleman, P., Doi, Y., Duband, L., Eggens,
  M., Evers, J., Funaki, I., Gao, J.~R., Giard, M., di~Giorgio, A.,
  Fern{\'{a}}ndez, L. M.~G., Griffin, M., Helmich, F.~P., Hijmering, R.,
  Huisman, R., Ishihara, D., Isobe, N., Jackson, B., Jacobs, H., Jellema, W.,
  Kamp, I., Kaneda, H., Kawada, M., Kemper, F., Kerschbaum, F., Khosropanah,
  P., Kohno, K., Kooijman, P.~P., Krause, O., van~der Kuur, J., Kwon, J.,
  Laauwen, W.~M., de~Lange, G., Larsson, B., van Loon, D., Madden, S.~C.,
  Matsuhara, H., Najarro, F., Nakagawa, T., Naylor, D., Ogawa, H., Onaka, T.,
  Oyabu, S., Poglitsch, A., Reveret, V., Rodriguez, L., Spinoglio, L., Sakon,
  I., Sato, Y., Shinozaki, K., Shipman, R., Sugita, H., Suzuki, T., van~der
  Tak, F. F.~S., Redondo, J.~T., Wada, T., Wang, S.~Y., Wafelbakker, C.~K., van
  Weers, H., Withington, S., Vandenbussche, B., Yamada, T., and Yamamura, I.,
  ``Spica—a large cryogenic infrared space telescope: Unveiling the obscured
  universe,'' {\em Publications of the Astronomical Society of Australia}~{\bf
  35} (2018).

\bibitem{Makiwa:13}
Makiwa, G., Naylor, D.~A., Ferlet, M., Salji, C., Swinyard, B., Polehampton,
  E., and van~der Wiel, M. H.~D., ``Beam profile for the {H}erschel-{SPIRE}
  {F}ourier {T}ransform {S}pectrometer,'' {\em Appl. Opt.}~{\bf 52},
  3864--3875 (Jun 2013).

\bibitem{Swinyard:14}
{Swinyard}, B.~M., {Polehampton}, E.~T., {Hopwood}, R., {Valtchanov}, I., {Lu},
  N., {Fulton}, T., {Benielli}, D., {Imhof}, P., {Marchili}, N., {Baluteau},
  J.~., {Bendo}, G.~J., {Ferlet}, M., {Griffin}, M.~J., {Lim}, T.~L., {Makiwa},
  G., {Naylor}, D.~A., {Orton}, G.~S., {Papageorgiou}, A., {Pearson}, C.~P.,
  {Schulz}, B., {Sidher}, S.~D., {Spencer}, L.~D., v.~d. {Wiel}, M. H.~D., and
  {Wu}, R., ``Calibration of the herschel spire fourier transform
  spectrometer,'' {\em Monthly Notices of the Royal Astronomical Society}~{\bf
  440}(4),  3658--3674 (2014).

\bibitem{Valtchanov:17}
Valtchanov, I., Hopwood, R., Bendo, G., Benson, C., Conversi, L., Fulton, T.,
  Griffin, M.~J., Joubaud, T., Lim, T., and Lu, N., ``Correcting the
  extended-source calibration for the herschel-spire fourier-transform
  spectrometer,'' {\em Monthly Notices of the Royal Astronomical Society}~{\bf
  475},  321–330 (Dec 2017).

\bibitem{Griffin:10}
{Griffin, M. J.}, {Abergel, A.}, {Abreu, A.}, {Ade, P. A. R.}, {Andr\'e, P.},
  {Augueres, J.-L.}, {Babbedge, T.}, {Bae, Y.}, {Baillie, T.}, {Baluteau,
  J.-P.}, {Barlow, M. J.}, {Bendo, G.}, {Benielli, D.}, {Bock, J. J.},
  {Bonhomme, P.}, {Brisbin, D.}, {Brockley-Blatt, C.}, {Caldwell, M.}, {Cara,
  C.}, {Castro-Rodriguez, N.}, {Cerulli, R.}, {Chanial, P.}, {Chen, S.},
  {Clark, E.}, {Clements, D. L.}, {Clerc, L.}, {Coker, J.}, {Communal, D.},
  {Conversi, L.}, {Cox, P.}, {Crumb, D.}, {Cunningham, C.}, {Daly, F.}, {Davis,
  G. R.}, {De Antoni, P.}, {Delderfield, J.}, {Devin, N.}, {Di Giorgio, A.},
  {Didschuns, I.}, {Dohlen, K.}, {Donati, M.}, {Dowell, A.}, {Dowell, C. D.},
  {Duband, L.}, {Dumaye, L.}, {Emery, R. J.}, {Ferlet, M.}, {Ferrand, D.},
  {Fontignie, J.}, {Fox, M.}, {Franceschini, A.}, {Frerking, M.}, {Fulton, T.},
  {Garcia, J.}, {Gastaud, R.}, {Gear, W. K.}, {Glenn, J.}, {Goizel, A.},
  {Griffin, D. K.}, {Grundy, T.}, {Guest, S.}, {Guillemet, L.}, {Hargrave, P.
  C.}, {Harwit, M.}, {Hastings, P.}, {Hatziminaoglou, E.}, {Herman, M.},
  {Hinde, B.}, {Hristov, V.}, {Huang, M.}, {Imhof, P.}, {Isaak, K. J.},
  {Israelsson, U.}, {Ivison, R. J.}, {Jennings, D.}, {Kiernan, B.}, {King, K.
  J.}, {Lange, A. E.}, {Latter, W.}, {Laurent, G.}, {Laurent, P.}, {Leeks, S.
  J.}, {Lellouch, E.}, {Levenson, L.}, {Li, B.}, {Li, J.}, {Lilienthal, J.},
  {Lim, T.}, {Liu, S. J.}, {Lu, N.}, {Madden, S.}, {Mainetti, G.}, {Marliani,
  P.}, {McKay, D.}, {Mercier, K.}, {Molinari, S.}, {Morris, H.}, {Moseley, H.},
  {Mulder, J.}, {Mur, M.}, {Naylor, D. A.}, {Nguyen, H.}, {O'Halloran, B.},
  {Oliver, S.}, {Olofsson, G.}, {Olofsson, H.-G.}, {Orfei, R.}, {Page, M. J.},
  {Pain, I.}, {Panuzzo, P.}, {Papageorgiou, A.}, {Parks, G.}, {Parr-Burman,
  P.}, {Pearce, A.}, {Pearson, C.}, {P\'erez-Fournon, I.}, {Pinsard, F.},
  {Pisano, G.}, {Podosek, J.}, {Pohlen, M.}, {Polehampton, E. T.}, {Pouliquen,
  D.}, {Rigopoulou, D.}, {Rizzo, D.}, {Roseboom, I. G.}, {Roussel, H.},
  {Rowan-Robinson, M.}, {Rownd, B.}, {Saraceno, P.}, {Sauvage, M.}, {Savage,
  R.}, {Savini, G.}, {Sawyer, E.}, {Scharmberg, C.}, {Schmitt, D.}, {Schneider,
  N.}, {Schulz, B.}, {Schwartz, A.}, {Shafer, R.}, {Shupe, D. L.}, {Sibthorpe,
  B.}, {Sidher, S.}, {Smith, A.}, {Smith, A. J.}, {Smith, D.}, {Spencer, L.},
  {Stobie, B.}, {Sudiwala, R.}, {Sukhatme, K.}, {Surace, C.}, {Stevens, J. A.},
  {Swinyard, B. M.}, {Trichas, M.}, {Tourette, T.}, {Triou, H.}, {Tseng, S.},
  {Tucker, C.}, {Turner, A.}, {Vaccari, M.}, {Valtchanov, I.}, {Vigroux, L.},
  {Virique, E.}, {Voellmer, G.}, {Walker, H.}, {Ward, R.}, {Waskett, T.},
  {Weilert, M.}, {Wesson, R.}, {White, G. J.}, {Whitehouse, N.}, {Wilson, C.
  D.}, {Winter, B.}, {Woodcraft, A. L.}, {Wright, G. S.}, {Xu, C. K.},
  {Zavagno, A.}, {Zemcov, M.}, {Zhang, L.}, and {Zonca, E.}, ``The
  herschel-spire instrument and its in-flight performance*,'' {\em A\&A}~{\bf
  518},  L3 (2010).

\bibitem{Griffin:03}
``Spire design description.'' European Space Agency website, 2022
  \url{https://www.cosmos.esa.int/documents/12133/1035800/SPIRE+Design+Description}.
\newblock (Accessed: 1 June 2022).

\bibitem{Mach:92}
Mach, L., ``Ueber einen interferenzrefraktor,'' {\em Zeitschrift für
  Instrumentenkunde} (12),  89–93 (1892).

\bibitem{Zehnder:91}
Zehnder, L., ``Ein neuer interferenzrefraktor,'' {\em Zeitschrift für
  Instrumentenkunde} (11),  275–285 (1891).

\bibitem{Michelson:90}
Michelson, A.~A., ``Measurement by light-waves,'' {\em Am. J. Sci}~{\bf 39},
  115 (1890).

\bibitem{Goldsmith:98}
{Goldsmith}, P.~F., ``Quasi-optical techniques,'' {\em Proceedings of the
  IEEE}~{\bf 80}(11),  1729--1747 (1992).

\bibitem{Wolf:07}
Wolf, E.,  [{\em Introduction to the Theory of Coherence and Polarization of
  Light}{\nolinebreak\hspace{0.1em}]}, Cambridge University Press, 1st~ed.
  (2007).

\bibitem{Hecht:99}
Hecht, E.,  [{\em Optics}{\nolinebreak\hspace{0.1em}]}, Pearson Education
  Limited, 5th~ed. (2017).

\bibitem{Sein:03}
Sein, E., Toulemont, Y., Safa, F., Duran, M., Deny, P., Chambure, D.,
  Passvogel, T., and Pilbratt, G., ``A 3.5 m diameter sic telescope for
  herschel mission,'' {\em Proc. SPIE}~{\bf 4850} (01 2003).

\bibitem{Chen:18}
Jiajun, C., {\em Modal Optical Studies of Multi-Moded Ultra-Low-Noise Detectors
  in Far-Infrared}, PhD thesis, University of Cambridge (2018).

\bibitem{Orfanidis:16}
Orfanidis, S.,  [{\em Electromagnetic Waves and
  Antennas}{\nolinebreak\hspace{0.1em}]}, Sophocles J. Orfanidis (2016).

\bibitem{Goodman:05}
Goodman, J.~W., ``Introduction to fourier optics,'' {\em Introduction to
  Fourier optics, 3rd ed., by JW Goodman. Englewood, CO: Roberts \& Co.
  Publishers, 2005}~{\bf 1} (2005).

\bibitem{Wolf:82}
Wolf, E., ``New theory of partial coherence in the space--frequency domain.
  part i: spectra and cross spectra of steady-state sources,'' {\em J. Opt.
  Soc. Am.}~{\bf 72},  343--351 (Mar 1982).

\bibitem{Withington:01}
Withington, S. and Yassin, G., ``Power coupled between partially coherent
  vector fields in different states of coherence,'' {\em J. Opt. Soc. Am.
  A}~{\bf 18},  3061--3071 (Dec 2001).

\bibitem{Withington:07}
Withington, S. and Saklatvala, G., ``Characterizing the behaviour of partially
  coherent detectors through spatio-temporal modes,'' {\em Journal of Optics A:
  Pure and Applied Optics}~{\bf 9},  626 (06 2007).

\bibitem{Kano:62}
Kano, Y. and Wolf, E., ``Temporal coherence of black body radiation,'' {\em
  Proceedings of the Physical Society}~{\bf 80},  1273--1276 (dec 1962).

\bibitem{Withington:04}
Withington, S., Hobson, M.~P., and Berry, R.~H., ``Representing the behavior of
  partially coherent optical systems by using overcomplete basis sets,'' {\em
  J. Opt. Soc. Am. A}~{\bf 21},  207--217 (Feb 2004).

\bibitem{Locke:06}
Spencer, L.~D., Naylor, D.~A., and Swinyard, B.~M., ``{A comparison of the
  theoretical and measured performance of the Herschel/SPIRE imaging Fourier
  transform spectrometer},'' in [{\em Space Telescopes and Instrumentation I:
  Optical, Infrared, and Millimeter}{\nolinebreak\hspace{0.1em}]},  Mather,
  J.~C., MacEwen, H.~A., and de~Graauw, M. W.~M., eds.,  {\bf 6265},  968 --
  979, International Society for Optics and Photonics, SPIE (2006).

\bibitem{Gibbs:89}
Gibbs, J.~W., ``Fourier's series,'' {\em Nature}~{\bf 59},  200 (1989).

\bibitem{Chattopadhyay:03}
Chattopadhyay, G., Glenn, J., Bock, J., Rownd, B., Caldwell, M., and Griffin,
  M., ``Feed horn coupled bolometer arrays for spire - design, simulations, and
  measurements,'' {\em IEEE Transactions on Microwave Theory and
  Techniques}~{\bf 51}(10),  2139--2146 (2003).

\bibitem{Ramo:12}
Ramo, S., Whinnery, J.~R., and Van~Duzer, T.,  [{\em
  Optics}{\nolinebreak\hspace{0.1em}]}, John Wiley \& Sons, Inc., 3th~ed.
  (2012).

\bibitem{Wu:13}
{Wu, R.}, {Polehampton, E. T.}, {Etxaluze, M.}, {Makiwa, G.}, {Naylor, D. A.},
  {Salji, C.}, {Swinyard, B. M.}, {Ferlet, M.}, {van der Wiel, M. H. D.},
  {Smith, A. J.}, {Fulton, T.}, {Griffin, M. J.}, {Baluteau, J.-P.}, {Benielli,
  D.}, {Glenn, J.}, {Hopwood, R.}, {Imhof, P.}, {Lim, T.}, {Lu, N.}, {Panuzzo,
  P.}, {Pearson, C.}, {Sidher, S.}, and {Valtchanov, I.}, ``Observing extended
  sources with the herschel spire fourier transform spectrometer,'' {\em
  A\&A}~{\bf 556},  A116 (2013).

\bibitem{Fulton:14}
Fulton, T., Hopwood, R., Baluteau, J.-P., Benielli, D., Imhof, P., Lim, T., Lu,
  N., Marchili, N., Naylor, D., Polehampton, Edward~Swinyard, B., and
  Valtchanov, I., ``Herschel spire fts relative spectral response
  calibration,'' {\em Experimental Astronomy}~{\bf 37},  381--395 (2014).

\bibitem{Hopwood:15}
Hopwood, R., Polehampton, E.~T., Valtchanov, I., Swinyard, B.~M., Fulton, T.,
  Lu, N., Marchili, N., van~der Wiel, M. H.~D., Benielli, D., Imhof, P.,
  Baluteau, J.-P., Pearson, C., Clements, D.~L., Griffin, M.~J., Lim, T.~L.,
  Makiwa, G., Naylor, D.~A., Noble, G., Puga, E., and Spencer, L.~D.,
  ``{Systematic characterization of the Herschel SPIRE Fourier Transform
  Spectrometer},'' {\em Monthly Notices of the Royal Astronomical Society}~{\bf
  449},  2274--2303 (03 2015).

\end{thebibliography}
\bibliographystyle{spiebib} 

\end{document}